\documentclass[a4paper,11pt]{article}
\pdfoutput=1 % if your are submitting a pdflatex (i.e. if you have
\usepackage{jcappub} % for details on the use of the package, please
\usepackage[T1]{fontenc} % if needed

\usepackage{graphics}
\usepackage{txfonts}
\usepackage{comment}

\renewcommand{\d}[1]{\ensuremath{\operatorname{d}\!{#1}}}

\newcommand{\infr}{<\! r}
\newcommand{\Nr}{N(<\! r)}

\newcommand{\sNr}{\mathcal{N}(<\! r)}
\newcommand{\sNrHat}[1]{\widehat{\mathcal{N}}(<\! r)}

\newcommand{\xiHat}[1]{\widehat\xi(r)}
\newcommand{\DHat}[1]{\widehat{D}_2(r)}
 \title{\boldmath A 14 $h^{-3}$ Gpc$^3$ study of cosmic homogeneity using BOSS DR12 quasar sample}

\author[a]{Pierre Laurent,}
\author[a,1]{Jean-Marc Le Goff,\note{Corresponding author.}}
\author[a]{Etienne Burtin,}
\author[b]{Jean-Christophe Hamilton,}
\author[c]{David W. Hogg,} 
\author[d]{Adam Myers,}
\author[b]{Pierros Ntelis,}
\author[e]{Isabelle P\^aris,}
\author[a]{James Rich,}
\author[b]{Eric Aubourg,}
\author[b,f]{Julian Bautista,}
\author[g]{Timoth\'ee Delubac,} % ??
\author[a]{H\'elion du Mas des Bourboux,}
\author[d]{Sarah Eftekharzadeh,}
\author[a]{Nathalie Palanque Delabrouille,}
\author[h]{Patrick Petitjean,}
\author[a,i]{Graziano Rossi,}
\author[j,k]{Donald P. Schneider,}
\author[a]{and Christophe Yeche}

\affiliation[a]{CEA, Centre de Saclay, IRFU/SPP,  F-91191 Gif-sur-Yvette, France}
\affiliation[b]{APC, Universit\'{e} Paris Diderot-Paris 7, CNRS/IN2P3, CEA, Observatoire de Paris,\\ 10, rue A. Domon \& L. Duquet,  Paris, France}
\affiliation[c]{Center for Cosmology and Particle Physics, New York University, 4 Washington Place, Meyer Hall of Physics, New York, NY 10003, USA}
\affiliation[d]{Department of Physics and Astronomy, University of Wyoming, Laramie, WY 82071, USA}
\affiliation[e]{Aix Marseille Université, CNRS, LAM (Laboratoire d'Astrophysique de Marseille) UMR 7326, 13388, Marseille, France}
\affiliation[f]{Department of Physics and Astronomy, University of Utah, Salt Lake City, UT 84112, USA.}
\affiliation[g]{Laboratoire d'astrophysique, Ecole Polytechnique F\'ed\'erale de Lausanne (EPFL), Observatoire de Sauverny,CH-1290 Versoix, Switzerland}
\affiliation[h]{Institut d'Astrophysique de Paris, CNRS-UPMC, UMR7095, \\ 98bis bd Arago, Paris, 75014 France}
\affiliation[i]{Department of Astronomy and Space Science, Sejong University, Seoul, 143-747, Korea}
\affiliation[j]{Department of Astronomy and Astrophysics, The Pennsylvania State University, University Park, PA 16802}
\affiliation[k]{Institute for Gravitation and the Cosmos, The Pennsylvania State University, University Park, PA 16802}

\emailAdd{jmlegoff@cea.fr}

\date{Received xx xx 2014 / accepted  xx xx 2014}

\abstract{
The BOSS quasar sample is used to study cosmic homogeneity with a 3D survey in the redshift range $2.2<z<2.8$. We measure the count-in-sphere, $\Nr$, i.e.~the average number of objects around a given object, and its logarithmic derivative, the fractal correlation dimension, $D_2(r)$. For a homogeneous distribution $\Nr \propto r^3$ and $D_2(r)=3$. Due to the uncertainty on tracer density evolution, 3D surveys  can only probe homogeneity up to a redshift dependence, i.e. they probe so-called ``spatial isotropy". 
Our data demonstrate spatial isotropy of the quasar distribution in the redshift range $2.2<z<2.8$ in a model-independent way, independent of any FLRW fiducial cosmology, resulting in  $3-\langle D_2 \rangle < 1.7 \times 10^{-3}$ (2 $\sigma$) over the range $250<r<1200 \, h^{-1}$Mpc for the quasar distribution.
If we assume that quasars do not have a bias much less than unity, this implies spatial isotropy of the matter distribution on large scales.
Then, combining with the Copernican principle, we finally get homogeneity of the matter distribution on large scales. 
Alternatively, using a flat $\Lambda$CDM fiducial cosmology with CMB-derived parameters, and measuring the quasar bias relative to this $\Lambda$CDM model, our data provide a consistency check of the model, in terms of how homogeneous the Universe is on different scales. $D_2(r)$ is found to be compatible with our $\Lambda$CDM model on the whole $10<r<1200 \, h^{-1}$Mpc range. For the matter distribution  we obtain $3-\langle D_2 \rangle < 5 \times 10^{-5}$ (2 $\sigma$) over the range $250<r<1200 \, h^{-1}$Mpc, consistent with homogeneity on large scales. 
}

\arxivnumber{1602.09010}

\begin{document}
\toccontinuoustrue
%\linenumbers
\maketitle
\flushbottom

%%%%%%%%%%%%%%%%%%%%%%%%%%%%%%%%%%%%%%%%%%%%%%%%%%%
\section{Introduction}

Modern cosmology relies on Cosmological Principle, i.e.~the assumption that the Universe is isotropic and homogeneous on large scales, up to small statistical fluctuations. Isotropy is tested at the $10^{-5}$ level using the cosmic microwave background \citep{Smoot+92} and additional tests are provided by the X-ray background \citep{Scharf+00} and the isotropy of radio galaxies \citep{Peebles93}. 
One should distinguish  ``spatial'' isotropy, i.e.~the assumption that $\rho(r,\theta_1,\phi_1)=\rho(r,\theta_2,\phi_2)$ for any $(r,\theta_1,\phi_1,\theta_2,\phi_2)$, and the isotropy in terms of the projected field, $\rho_{\rm proj}(\theta,\phi)=\int\rho(r,\theta,\phi)W(r)dr$, where $W(r)$ is the window function, i.e.~the assumption that $\rho_{\rm proj}(\theta_1,\phi_1)=\rho_{\rm proj}(\theta_2,\phi_2)$ for any $(\theta_1,\phi_1,\theta_2,\phi_2)$. 
The combination of spatial isotropy and the Copernican principle, which states that we are not in a special location in the Universe, implies homogeneity~ \citep{Straumann74,Peacock99,Clarkson12}.
This is not true for ``projected" isotropy, as explicitly shown by \citet{Durrer+97} 
who provide an example of a fractal set 
that has projected isotropy for an observer located anywhere on the fractal set, while the fractal set is of course not homogeneous on any scale.

CMB data indicate isotropy at one single given distance, while X-ray background and radio-galaxy data test projected isotropy. This is not enough to ensure homogeneity when combined with the Copernican principle. It is then desirable to test homogeneity. 
The transition from inhomogeneity on small scales to statistical homogeneity on large scales should also be studied and compared to model predictions.

This investigation has been performed with spectroscopic galaxy surveys, which provide a 3D map of the distribution of the galaxies. Some studies indeed found a transition to homogeneity at a scale between 70 and 150 $h^{-1}$Mpc \cite{martinez+94,guzzo97,martinez+98,scaramella+98,amendola+99,Pan+00,kurokawa+01,yadav+05,sarkar+09}, 
but other investigations failed to find this transition \citep{Coleman+92,Pietronero+97,Labini+98,Joyce+99,Labini+09,Labini11}. A definitive answer requires large and dense surveys with uniform selection and sufficiently simple geometry. Recent results obtained using the 
SDSS~II~\citep{Hogg+05,Nadathur13} and WiggleZ~\citep{Scrimgeour+12} data, which all found a transition to homogeneity, are therefore particularly relevant.

Galaxy surveys actually only measure density on our past light-cone and not inside of it. Star formation history was used in an attempt to overcome this limitation and study homogeneity inside our past light-cone~\cite{Hoyle+13} but this is probably model dependent.
Homogeneity was also positively tested using a combination of secondary CMB probes, including integrated Sachs-Wolfe, kinetic Sunyaev-Zel'dovich and Rees-Sciama effects, and CMB lensing~\cite{Zibin+14}. These techniques allow homogeneity to be tested at much higher redshifts than galaxy surveys, up to recombination at $z\approx1100$.

In this paper we use the quasar sample of the Baryon Oscillation Spectroscopic Survey (BOSS) \citep{bossoverview} to study cosmic homogeneity with a 3D survey. A similar study is ongoing using the BOSS luminous-red-galaxy sample \citep{Ntelis+16}. 
Quasars are rare objects and large enough samples to allow for detailed clustering studies
\cite{Croom+01,Porciani+04,Croom+05,Hennawi+06,Myers+06,PorcianiNorberg06,Myers+07a,Myers+07b,daAngela+08,Cro09,Shen+09,Ross+09} have only appeared recently. 
A  first study of cosmic homogeneity using a quasar sample~\cite{Nadathur13} was published in 2013. It used the SDSS-II sample
and included 18,722 quasars over the range $1.0 \le z \le 1.8$.
Our final sample includes 38,382 quasars over the range $2.2<z<2.8$. It is obtained using an algorithm specially designed to produce a uniform selection and is therefore more adapted to study homogeneity than the SDSS-II sample.
The volume of the part of the BOSS survey that we use, 14 $h^{-3}$ Gpc$^3$, is comparable to what was used in the study with SDSS-II quasars~\citep{Nadathur13}, 16 $h^{-3}$ Gpc$^3$, 
and much larger than for studies with galaxies: 0.25 $h^{-3}$ Gpc$^3$ for SDSS~II~\citep{Hogg+05} and $\sim$1 $h^{-3}$ Gpc$^3$ for WiggleZ \citep{Scrimgeour+12}.

We use Planck 2013 + WMAP9 parameters~\cite{Planck13} as our fiducial flat $\Lambda$CDM cosmology, namely $h=0.6704$, $\Omega_m=0.3183$, $\Omega_b h^2=0.022032$, $n_s=0.9619$ and $\sigma_8=0.8347$. 
Throughout this paper, magnitudes use the asinh scale at low flux levels, as described by~\citet{Lupton+99}.
The paper is organized as follows. Section 2 describes the quasar data sample used in this analysis, section 3 introduces the observables used to quantify the cosmic homogeneity and describes the analysis, section 4 discusses the effects of systematics, section 5 presents the results, and conclusions are drawn in section 6.

%%%%%%%%%%%%%%%%%%%%%%%%%%%%%%%%%%%%%%%%%%%%%%%%%%%
\section{Observational sample}
\label{sec:setup}

\subsection{BOSS survey}

The BOSS project of the Sloan Digital Sky Survey (SDSS-III) \citep{eisenstein11} was designed to obtain the spectra of over $\sim1.6\times10^6$ luminous galaxies  and  $\sim150,000$ quasars. The project uses upgraded versions of the SDSS spectrographs \citep{bossspectrometer} mounted on the Sloan 2.5-meter telescope \citep{gunn06} at Apache Point Observatory, New Mexico. 

An aluminum plate is set at the focal plane of the telescope with a $3^{\circ}$ diameter field-of-view. Holes are drilled in the plate, corresponding  to 1000 targets, i.e., objects to be observed with one of the two spectrographs. An optical fiber is plugged to each hole and sent to the spectrographs. The minimum distance between two fibers on the same plate corresponds to 62'' on the sky, which results in some collisions between targets. It may, however, be possible to observe both colliding targets if they are in the overlap region between two or more plates.
 
The SDSS photometric images in 5 bands ({\it u,g,r,i,z}) \citep{Fukugita+96} are used to define the targets. Selecting quasars in the redshift range $z\simeq 2$--3 is difficult due to a background of objects with similar colors, namely metal-poor A and F stars, faint lower redshift quasars and compact galaxies \citep[e.g.,][]{Fan99,Ric01}. 
In addition, BOSS operated close to the detection limits of the SDSS photometry, where uncertainties on flux measurements cause objects to scatter substantially in color space.

\subsection{Quasar selection}

The study of cosmic homogeneity requires a uniform selection across the sky, while BOSS is selecting quasars mainly for a Lyman-$\alpha$ Forest survey, which requires as many quasars as possible in a given redshift range but not necessarily with a uniform selection. The solution to reconcile these competing requirements was to define a CORE sample with uniform selection and a BONUS sample that aims at selecting as many quasars as possible using all information available in each patch of the sky. 
The CORE sample produces, on average, 20 targets per deg$^2$.
In each plate, these CORE targets are completed with BONUS targets up to an allowed budget of 40 targets per deg$^2$.
The CORE  sample is selected using the extreme deconvolution (XD) algorithm,\footnote{XD \citep{Bovy+09} is a method to describe the underlying distribution function of a series of points in parameter space (e.g., quasars in color space) by modeling that distribution as a sum of Gaussians convolved with measurement errors.} which is applied in BOSS to model the distributions of quasars and stars in flux space, and hence to separate quasar targets from stellar contaminants \citep[XDQSO;][]{Bovy+11}.

In this paper, we consider as quasar targets all point sources in SDSS imaging 
up to the magnitude limit of BOSS quasar target selection, $g\leq22.0$ or $r\leq 21.85$, 
that have an XDQSO probability above a threshold of $0.424$ \citep{ross12}.
We use data from the final DR12 data release of SDSS-III~\citep{DR11-12}. 
Data are split into groups corresponding to a given version of the targeting algorithm, which are called chunks. 
During the first year of operation a different algorithm was used to define the CORE sample. Therefore some XDQSO targets are not included in our first-year sample and these data cannot easily be used to study homogeneity. This constraint removes all chunks up to chunk 11.
For chunks 12 and 13, in case of collision, the CMASS galaxy targets had priority over the CORE targets~\citep{ross12}. Chunks 12 and 13 would require a special treatment involving masks around each galaxy target. Since our sample provides in any case a good statistical accuracy for the study of homogeneity, we prefer to limit ourselves to data that were taken in the same conditions and can be analyzed in a consistent manner. 
Removing these data, we are left with a small irregular patch of 1335 deg$^2$ in the South Galactic Cap (SGC). 
This patch is not optimal to study homogeneity, so we decided to remove all SGC data. 
Nevertheless, if we compute the fractal correlation dimension in the SGC, it is statistically compatible with that of the NGC, but with much larger error bars.

Our final sample consists of North Galactic Cap (NGC) data from chunks 14 and above. 
Figure~\ref{fig:footprint} presents the resulting footprint, which covers 5983 deg$^2$ and includes 112,193 CORE targets.
This footprint has an average of 20.2 targets per deg$^2$, out of which 13.9 actually have spectra consistent with them being a quasar. 
The BOSS pipeline~\citep{Bolton+12} identifies all quasar targets and provides their redshifts. All spectra are then visually inspected~\cite{Paris+12,Paris+14}, to correct for misidentifications or inaccurate redshift determinations.
The redshift is determined using the Mg II emission line if it is present in the spectrum, clearly detected, and not affected by sky subtraction. In other cases the redshift is estimated using the position of the red wing of the C IV emission line~\cite{Paris+12}. 
The uncertainty on the redshift determination is $\sigma_z \approx 0.001$ at one standard deviation.\footnote{
\citet{Paris+12} gives $\Delta z \approx 0.003$, but this is actually a 3 $\sigma$ error.
}
Figure~\ref{fig:nz} presents the redshift distribution of the measured CORE targets. We limit the analysis to the range $2.2<z<2.8$, where the distribution is significant. Table~\ref{tab:targets} lists some statistics on CORE targets.

%%%%%%%%%%%%%%%%%%%%%%%%%%%%%%%%%%%%%%%%%%%%%%%%%%%
\section{Analysis}
\label{sec:analysis}

\subsection{Survey completeness}

\begin{figure}[t]
\begin{center}
\epsfig{figure=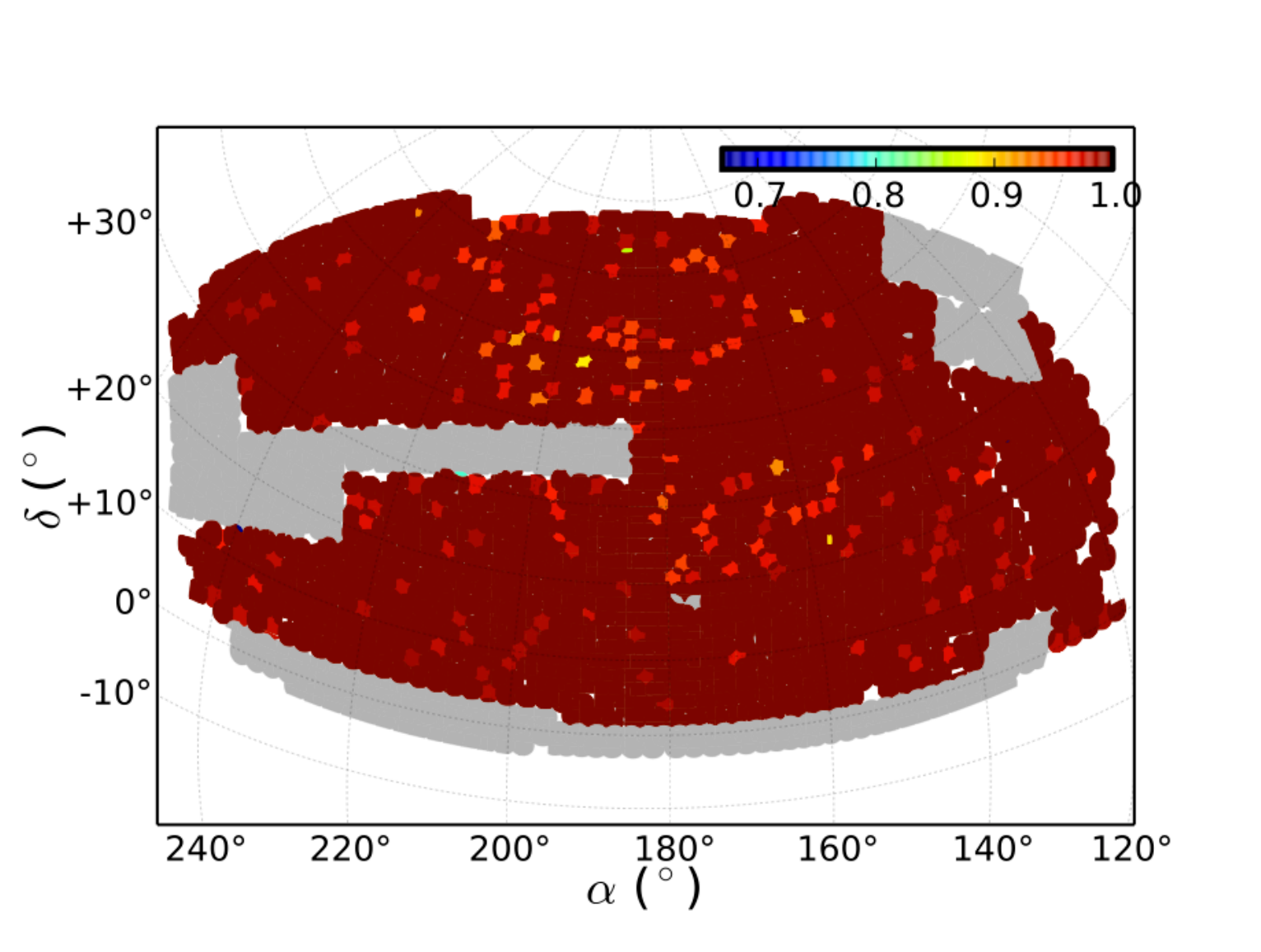 ,width = 10cm} 
\caption{\it  Angular distribution of the selected data (5983 deg$^2$). 
The color scale indicates the survey completeness in each polygon.
The grey area represents the full BOSS NGC footprint. }
\label{fig:footprint}
\end{center}
\end{figure}

The Mangle software \citep{Blanton+03,Tegmark+04,mangle} is used to define the geometry of the survey and to apply masks to remove bad areas. 
These bad areas include regions around bright stars and other bright objects, the center of each tile, which is covered by the centerpost of the cartridge, and bad photometric fields.
This masking is done using polygons, which are defined as  intersections of an arbitrary number of spherical caps on the celestial sphere. 

In our sample, in case of fiber collision, the CORE targets have the highest priority. However, there are 772 CORE targets (0.7\%),  which could not be observed due to a collision with another CORE target. In these cases, we follow \citet{anderson12} and assign a double weight to the colliding quasar that received a fiber. 
Since only 70\% of the targets actually are quasars, the weight should rather be smaller than two. 
But this is irrelevant to us since, after double weighting the colliding quasars, $\sNr$ varies around the uncorrected one with no particular trend and by less than half the statistical error on scales larger than 1 $h^{-1}$Mpc.

Fibers cannot be allocated to all targets due to the lack of available fibers on a given plate, 
and the targets that are not given a fiber are randomly selected. 
This effect is taken into account by defining a survey completeness,  
\begin{equation}
C=\frac{N_{obs}+N_{coll}}{N_{target}-N_{known}} \; ,
\end{equation}
in each polygon. Here $N_{obs}$ is the number of targets that were allocated a fiber and then observed, $N_{coll}$ is the number of targets that were not observed due to a collision with another target, $N_{target}$ is the total number of targets and $N_{known}$ refers to $z<2.15$ quasar targets that
had a spectrum measured previously and are not re-observed by BOSS. 
$N_{known}$ only includes known quasars with $z<2.15$ because known quasars at higher redshifts are re-observed by BOSS in order to improve the signal-to-noise ratio for the Lyman$-\alpha$ forest analysis. Therefore these quasars are treated as if they were not previously known.
The position of the collided targets is correlated with the observed targets, so collisions are accounted for by upweighthing, as explained above, rather than completeness.
Therefore $N_{coll}$ is added to the numerator of the completeness. 

Finally, $N_{obs}$ includes objects with a spectrum but that failed to be identified or that could not be ascribed a redshift. These objects are very unlikely to be quasars, so they do not produce any loss in our sample and we do not make any correction for them. 
With the CORE plus BONUS strategy, there are only 249 not-collided CORE targets that were not allocated a fiber, as can be inferred from table~\ref{tab:targets}. 	
The completeness is unity in most polygons, as illustrated in figure~\ref{fig:footprint}, and the average completeness is 0.998.

As discussed in section~\ref{sec:syst}, we apply a cut on the apparent {\it i}-band magnitude, $m_i < 21.3$, in order to mitigate the effects of systematics. 
In contrast with \citet{white12}, we do not cut on the absolute magnitude. Our focus is not on defining a sample of quasars with the same properties, so we retain all to improve the statistics for the cosmic homogeneity study. 

\begin{figure}[t]
\begin{center}
\epsfig{figure= 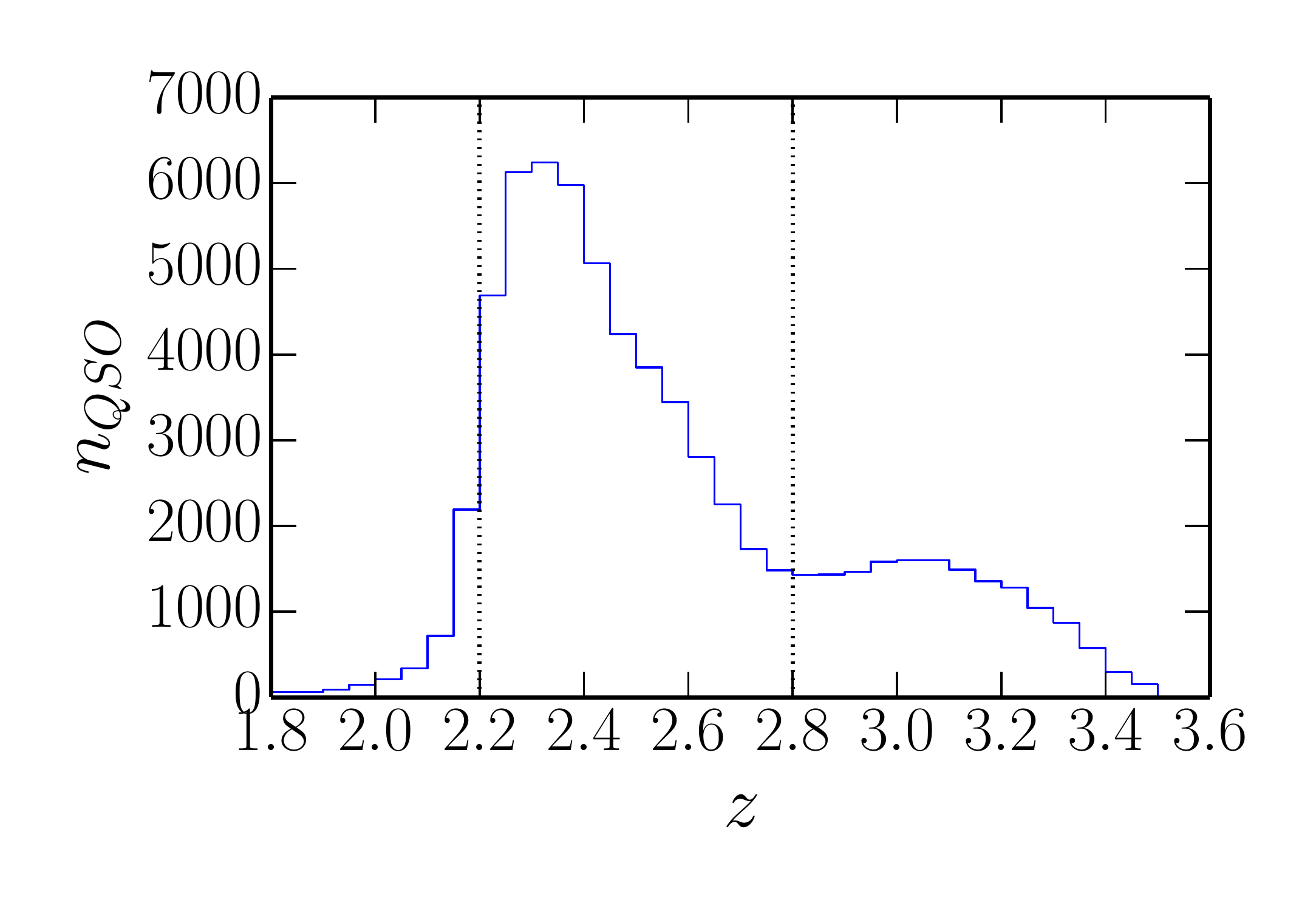,width = 8cm} 
\caption{\it  Redshift distribution of the measured CORE quasars (number of quasars in $\Delta z =0.05$ intervals). The two vertical lines define the sample selected for the analysis (47,858 quasars out of 77,203).}
\label{fig:nz}
\end{center}
\end{figure}

\begin{table}
\centering
\begin{tabular}{lr} \hline \hline
Targets in footprint & 112,193\\
Targets with a fiber & 111,172\\
Collisions between CORE & 772 \\
Quasars		& 77,203 \\
$2.2<z<2.8$ & 47,858\\
$m_i<21.3$ & 38,382\\	
\hline
\end{tabular}
\caption{Statistics of CORE targets.}
\label{tab:targets}
\end{table}

%--------------------------------------------------------------------------------------------------
\subsection{Observables}
\label{sec:obs}

In order to check for cosmic homogeneity it is important not to use a statistic that implicitly relies on the assumption of homogeneity. The evaluation of the correlation function, for instance, requires a mean density, which is only defined for a homogeneous sample. We use two related statistics, the counts-in-sphere $\Nr$, which is the average number of quasars in a sphere of radius $r$ around a given quasar, and the fractal correlation dimension,
\begin{equation}
D_2(r) \equiv \frac{\d \, \ln \Nr}{\d \,\ln r} \; .
\end{equation}

If the distribution is homogeneous, then $\Nr \propto r^3$ and $D_2(r)=3$. 
If the distribution is fractal, $D_2$ is a measure of the fractal dimension. 
However, this statement is only true if the survey itself does not introduce inhomogeneities, which it actually does due to the survey geometry and the completeness.
Following \citet{Scrimgeour+12} we introduce $\sNr$, which is the ratio of $\Nr$ to the same quantity for a random homogeneous sample. This random sample is Poisson distributed on the sky with the same geometry and completeness as the measured quasars. It also has the same redshift distribution.
For a homogeneous distribution, the scaled counts-in-sphere, $\sNr$, scales as $r^0$ and  $D_2$ is now defined as
\begin{equation}
\label{eq:d2}
D_2(r) \equiv \frac{\d\, \ln \sNr}{\d\, \ln r} +3 \; .
\end{equation}

We use the {\it ransack} code from Mangle to generate a catalogue of 1,000,000 random objects over the selected footprint: the number of random objects generated in each polygon is directly proportional to its area times its completeness.
The same masks, as applied to the data, are then applied to the random catalogue,
which reduces the number of random objects to 950,000.
This is 25 times the number of observed quasars in the same area. 
Since the statistical error declines as the square root of the number of pairs,
the statistical error related to the random catalogue is 25 times smaller than that of the data, which is negligible.
Each object in the catalogue is then assigned a redshift in the range $2.2 < z < 2.8$ by drawing randomly according to the measured redshift distribution $n(z)$. 
We are then insensitive to a possible isotropic variation of density with redshift, $\rho=\rho(r)$.
This is a general issue for all 3D surveys, they cannot, for example, falsify models where we live in the center of a spherical void.

Using the data and random catalogues, the following quantities are defined: $DD(r)$ is the density of pairs of objects at a distance $r$, normalized to the total number of pairs, $n(n-1)/2$; $RR(r)$ is the same quantity for the random catalogue and $DR(r)$ is the normalized density of pairs with one data object and one random object. The random catalogue takes the completeness and geometry into account.

The most straightforward estimators for the scaled count-in-sphere are 
$\left. \int_0^r DD(s) \d s \, \middle/  \int_0^r RR(s) \d s \right.$ 
or alternatively $\left. \int_0^r DD(s) \d s \, \middle/  \int_0^r DR(s) \d s \right.$, used by  \citet{Scrimgeour+12}.
We use a more sophisticated estimator, namely
\begin{equation}
\label{eq:nr-LS}
\sNrHat{LS}= 1+ \frac{\int_0^r \Bigl[DD(s)-2DR(s)+RR(s)\Bigr]\d s}{\int_0^r RR(s)\d s}\; .
\end{equation}
Appendix \ref{add:LS} explains how this estimator is obtained from the \citet{LandySzalay93} optimal estimator
of the correlation function, $\xiHat{LS}=(DD-2DR+RR)/RR$.
We will therefore designate eq.~\ref{eq:nr-LS} as the Landy-Szalay (LS) estimator for $\sNr$, in contrast with the $DD/RR$ and $DD/DR$ estimators aforementioned. 

In clustering studies, pairs of tracers are usually weighted according to the FKP weight \citep{FKP94} in order to improve the statistical accuracy: $w_{\rm FKP}= 1/(1+nP(k_0))$, where $n$ is the tracer density and $P(k_0)$ is the power spectrum at the relevant scale $k_0$.
A scale of 40 $h^{-1}$Mpc corresponds to $k \approx 0.16$ $h^{-1}$Mpc. Our quasar sample is sparse, $nP(k=0.16) < 0.015$ in all $z$ bins, and the weight is always larger than 0.985. The relative gain on the statistical error, $\sqrt{1+\sigma_w^2/\overline w^2}$, is on the order of $10^{-5}$. The gain is even smaller at scales larger than 40 $h^{-1}$Mpc.
This improvement is completely negligible and we do not use the FKP weight.  

The measurements are performed using quasars, which are biased tracers of the underlying total matter distribution, with a bias $b_Q$.
In order to evaluate $\sNr$ for the underlying matter distribution, the expression in eq.~\ref{eq:nr-LS} must be corrected for bias: 
\begin{equation}
\label{Eq:debias}
\sNr-1=\frac{ \mathcal{N}_Q(\infr)-1}{b_Q^2} \;.
\end{equation}
Once $\sNr$ has been corrected for bias, the correlation dimension $D_2$ is calculated as in eq.~\ref{eq:d2}. 
In the limit $|D_2-3| \ll 1$, eq.~\ref{Eq:debias} gives
\begin{equation}
\label{Eq:debiasD2}
D_2(r)-3=\frac{ D_{2,Q}(r)-3}{b_Q^2} \;.
\end{equation}
The bias is measured by fitting the two-point correlation function of our sample, $\xi(r)$, with a $\Lambda$CDM model. 

%--------------------------------------------------------------------------------------------------
\subsection{$\Lambda$CDM prediction for $P(k)$, $\xi(r)$, $\sNr$ and $D_2(r)$ }
\label{section:model}

Since the distances to the quasars are obtained from their redshifts, the peculiar velocities of quasars along the lines of sight distort the measured distribution of quasars.
This redshift-space distortion has two components which affect the matter power spectrum $P_{m}(k)$.  On large scales, there is a linear effect arising from the infall of quasars towards local over-densities~\citep{kaiser87}. The resulting quasar power spectrum in redshift space is :
$b_Q^2 \; (1+\beta\mu^2)^2\,P_{m}(k) \, ,$ 
where $\beta=f/b_Q$ and $f\approx\Omega_m^{0.55}(z)$ is the growth rate of structures~\citep{LinderCahn07}.
At small scales ($<15$ Mpc) we have the ``finger-of-God" effect ; the distortions become scale dependent and can be modeled using a {\it streaming} model~\citep{Peebles93}. This leads to: 
\begin{equation}
\label{Eq:Pmodel}
P_{Q}(k,\mu)=b^2\; (1+\beta\mu^2)^2\;\frac{1}{1+(k\,\sigma_{\! p}\,\mu)^2}\; P_{m}(k) \;,
\end{equation}
where $\sigma_p$ is the pair-wise velocity dispersion along the line of sight. 
This velocity dispersion depends on the mass of the dark matter halos inhabited by the quasars.
At our redshift, $z_{\rm eff} = 2.39$, an average halo mass $\langle M\rangle\approx2\times10^{12}\,M_{\odot}$ was inferred from the data using analytical models~\citep{Ross+09,white12,Eftekharzadeh+15}. The MultiDark Run N-body  simulations~\cite{Prada+12} indicate that the velocity dispersion for such haloes is $\sigma_p\approx270\,\mathrm{km.s^{-1}}$. The error on the redshift measurement, $\sigma_z \approx 0.001$ (see end of section~\ref{sec:setup}), is equivalent to an additional pair-wise velocity dispersion component of $125\,\mathrm{km.s^{-1}}$. Adding the two components in quadrature leads to an expected value of $\sigma_p\approx300\,\mathrm{km.s^{-1}}$.

The model for $P(k)$ is transformed into a model for $\xi(r)$ using a Fast Fourier Transform (FFT) and the parameters of the model (bias and velocity dispersion) are fitted to the correlation function.
We then have 
\begin{equation}
\sNr = \frac{\int_0^r [1+\xi(s)]s^2\d s }{ r^3/3}
\end{equation}
and $D_2(r)$ is found using eq.~\ref{eq:d2}.

%--------------------------------------------------------------------------------------------------
\subsection{Covariance matrix}
\label{sec:bootstrap}

Since mock catalogues do not exist for the BOSS quasar sample, we use bootstrap resampling \citep{EfronGong83} to estimate the covariance matrix of $\xi(r)$, $\sNr$ and $D_2(r)$. 
We split our footprint into 201 cells with identical volumes. 
For each cell, we  select 100 of the 200 other cells, and count the numbers of pairs $DD$, $DR$ and $RR$. 
We include both pairs with the two quasars in the considered cell
and pairs with one quasar in the considered cell and one quasar in one of the 100 selected cells. 
The selection of the 100 cells is done randomly but with the constraint that each pair of cells is selected exactly once. 
Drawing 201 cells {\it with replacement} out of the 201 cells provides a bootstrap realization.
We sum the numbers of pairs $DD$, $DR$ and $RR$ for the 201 cells and compute the ``measurements'', $\xi(r)$, $\sNr$ and $D_2(r)$ for this realization.
The covariance of the measurements for 1000 such bootstrap realizations provides an estimate of the covariance matrices.

The bootstrap procedure ignores cosmic variance, but this is not a serious problem in our case since, due to low density, shot noise strongly dominates.
In addition there is a small correlation between the results in neighboring cells. This feature does not fit into the conditions of applicability of the bootstrap procedure, but this is a small effect.

\begin{figure}[htb]
\begin{center}
\epsfig{figure= 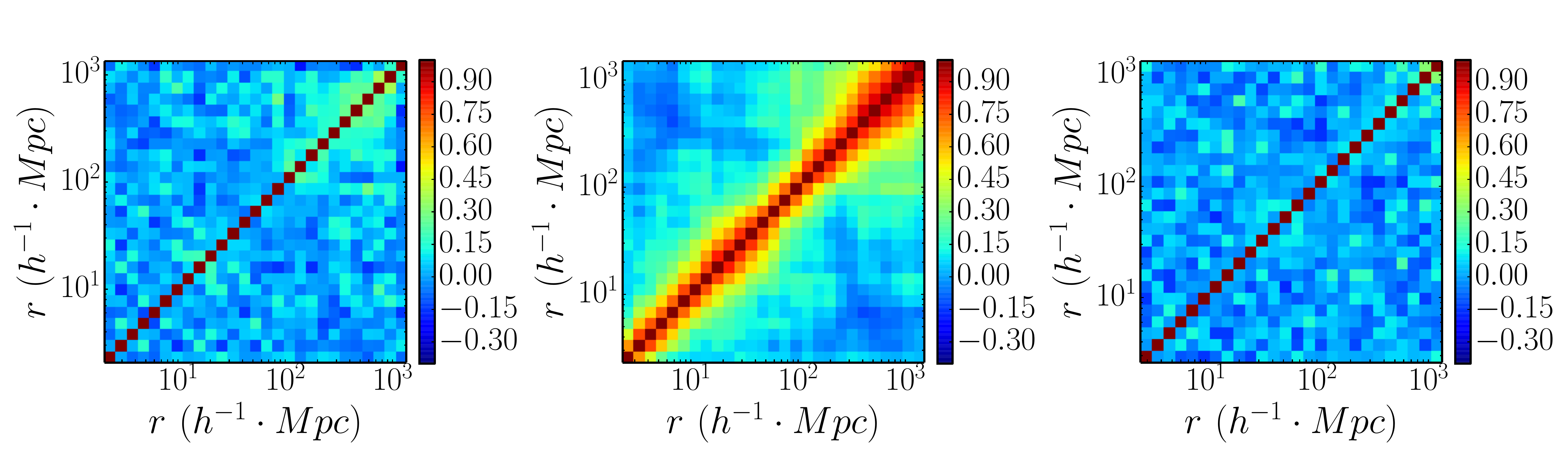,width = 15.cm} 
\caption{\it  The correlation matrix of $\xi(r)$, $\sNr$ and $D_2(r)$ from left to right.
$D_2(r)$ is essentially uncorrelated. It is thus a more appropriate observable than $\sNr$, which is significantly correlated.
}
\label{fig:corrmatrix}
\end{center}
\end{figure}

Figure~\ref{fig:corrmatrix} presents the resulting correlation matrix for $\xi$, $\sNr$ and $D_2(r)$.
The off-diagonal elements are quite small for $\xi(r)$, as expected for a shot-noise limited survey. 
A significant correlation is present for $\sNr$ since $\int_0^r DD(s) \d s$ is strongly correlated between two close values of $r$. The correlation matrix of $D_2(r)$ is strongly dominated by the diagonal, as for $\xi(r)$.

Varying the number of bootstrap cells from 25 to 201 does not change the general shape of the correlation matrices.
 It does, however, change individual elements of the diagonal of the covariance matrices by up to 20\%, 
but there is no particular trend, and this simply illustrates the noisiness of the covariance estimator.

\begin{figure}[t]
\begin{center}
\epsfig{figure= 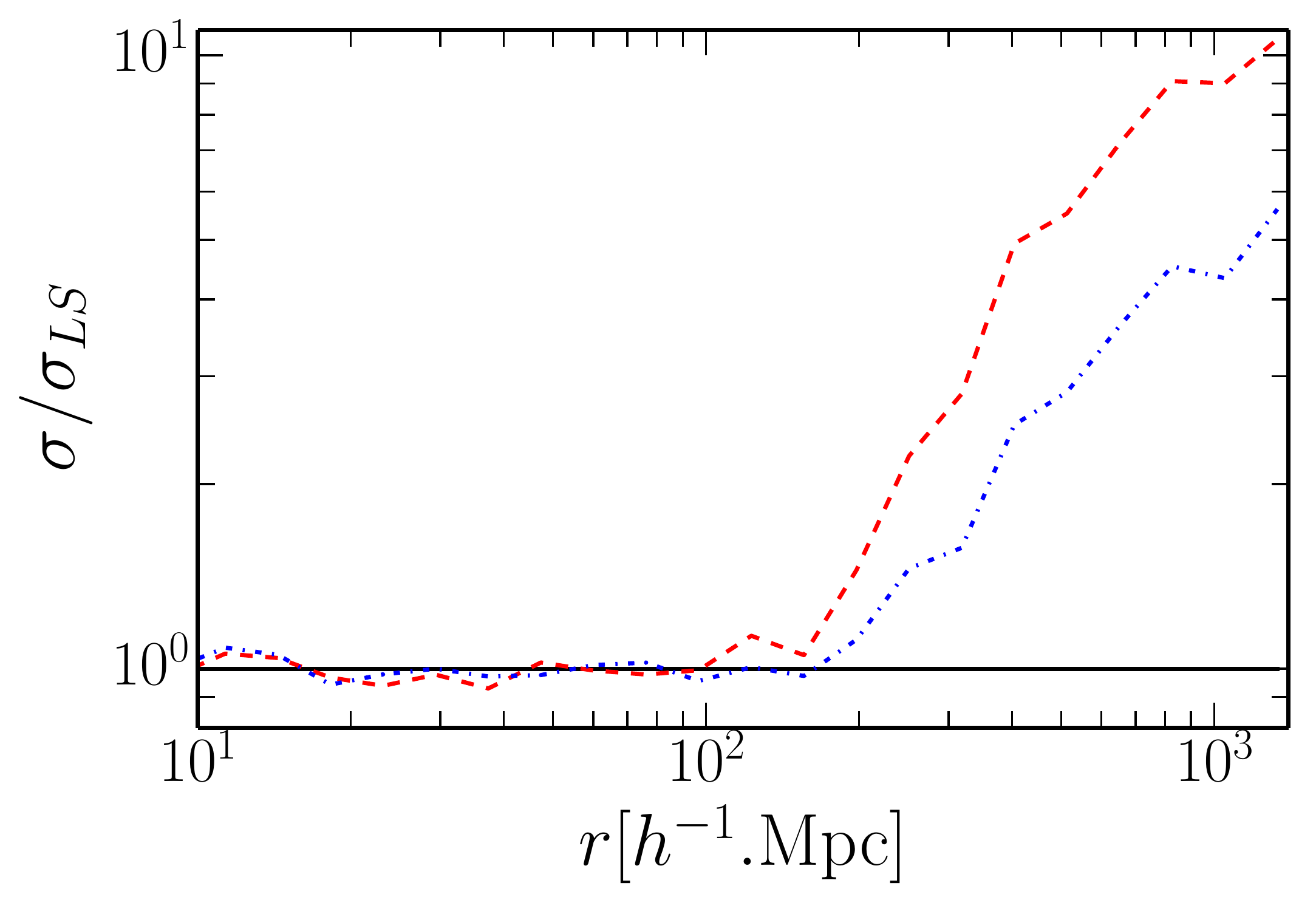,width = 8cm} 
\caption{\it  Ratio of the bootstrap error on $D_2(r)$ for the $DD/RR$ (red dashed line) and $DD/DR$ (blue dotted line) estimators to the LS estimator.
For $r>100 h^{-1}$ Mpc, there is a very significant gain in statistical accuracy with the LS estimator, up to a factor 10. 
}
\label{fig:error_ratio}
\end{center}
\end{figure}

Figure~\ref{fig:error_ratio} presents the ratio of the errors for the $DD/RR$  and $DD/DR$ estimators to the LS estimator.    %  <==  no PH 
This ratio is estimated with the bootstrap procedure for the case of $D_2(r)$. There is probably a contribution to the error in addition to the Poisson error, present for the $DD/RR$ and $DD/DR$ estimators and not for the LS estimator. On small scales the Poisson error dominates and the three estimators have the same error. On large scales there are many pairs, the Poisson error becomes small and there is a large gain with the LS estimator. This behavior is similar to what is described by \citet{LandySzalay93} for $\hat\xi(\theta)$.

%%%%%%%%%%%%%%%%%%%%%%%%%%%%%%%%%%%%%%%%%%%%%%%%%%%
\section{Effects of systematics}

\subsection{Data-related effects}
\label{sec:syst}
Investigating homogeneity requires a homogeneous target selection, which is difficult to achieve in our redshift range, as explained in section 2. The dependence of target-selection completeness on redshift is accounted for by the use of a random catalogue in our estimators, at least in the limit that it is independent of angular position. But a dependence of the completeness with angular position will bias the measurement. This is the main source of systematics. Indeed, errors on the angular position of the quasars are negligible (of the order of kpc), and those on the redshift, $\sigma_z \approx 0.001$ at one standard deviation (i.e.,~$\approx$~1 $h^{-1}$Mpc), were included 
in the model of the correlation function (see section \ref{section:model}).
Finally, it was previously mentioned that there are few collisions and they have a negligible effect.

\begin{figure}[h]
\begin{center}
\epsfig{figure=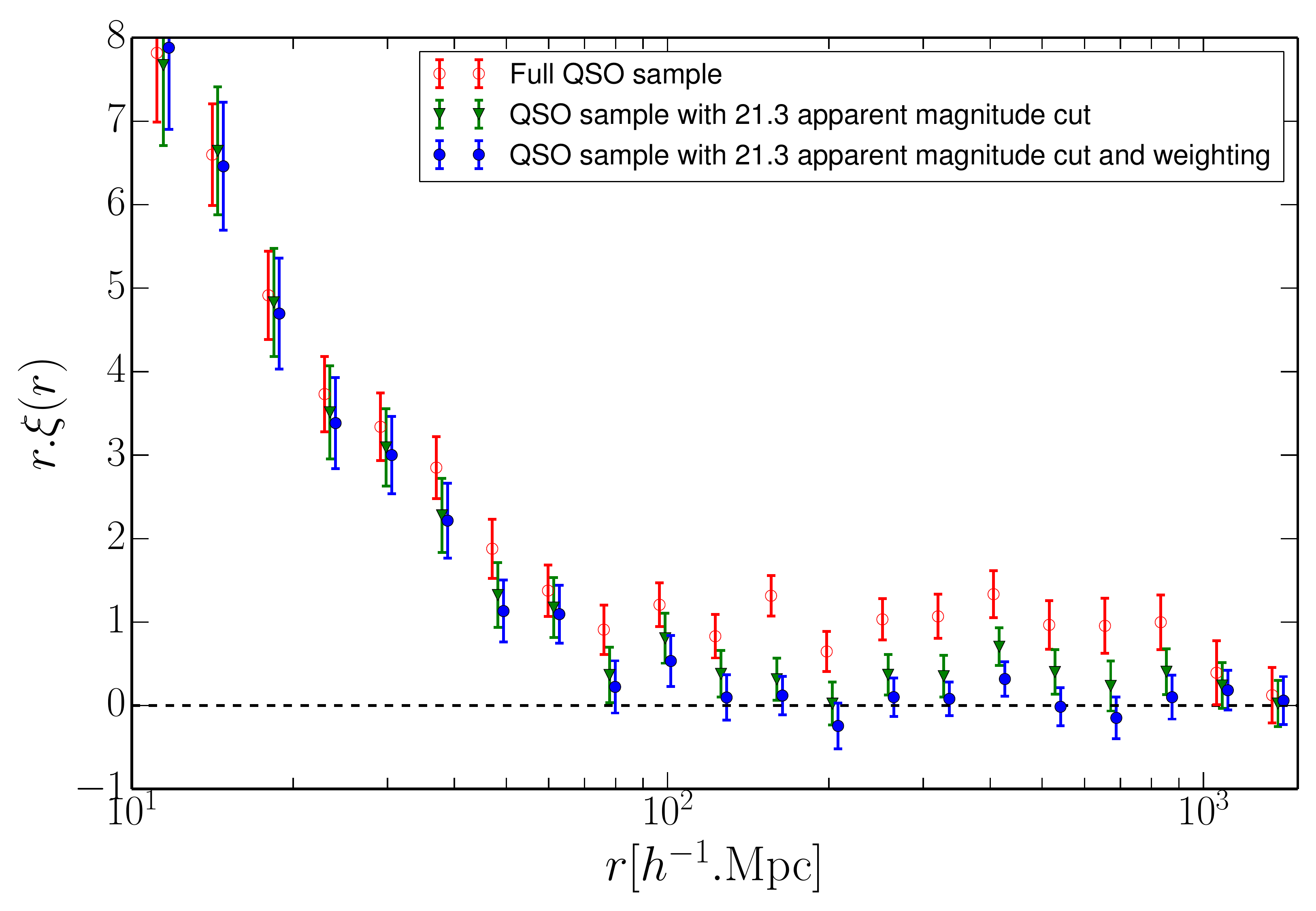,width = 12cm} 
\caption{\it The function $r.\xi(r)$ obtained with the LS estimator for our quasar sample without cut (red open circles), after a 21.3 magnitude cut (green triangles) and after additional weighting (blue circles). The same blue points appear in the left panel of figure~\ref{fig:xi}, where they can be compared to the prediction of our $\Lambda$CDM model.}
\label{fig:mag_cut}
\end{center}
\end{figure}

Quasar targets are selected in the SDSS photometric survey to a maximum limit $g \le 22$ or $r \le 21.85$~\cite{ross12},
while the average 5-$\sigma$ detection-limits of the survey for point sources are $g=23.1$ and $r= 22.7$. 
These detection limits vary with observational conditions such as seeing, airmass, Galactic extinction or sky brightness, by up to $\pm0.8$ magnitude. Some regions of the survey may then have targets with a magnitude just below the detection limit. Such targets have lower signal-to-noise ratio for their photometric fluxes and their intrinsic XDQSO probability is more likely to move below threshold.
In addition faint quasars are more sensitive to blending, i.e.,~biases on their flux measurement due to the proximity of another object. 
We therefore exclude quasars with large apparent magnitudes. In order to be homogeneous, this procedure is done on magnitude that has been corrected for Galactic extinction.
We chose to apply the cut in the $i$ band since there are already limits in the target selection in the $g$ and $r$ bands.
Starting from no cut and tightening the cut significantly changes the resulting correlation function, confirming that faint quasars are affected by the effects of systematics. When the threshold reaches $m_i = 21.3$, the correlation function stabilizes, providing a result less sensitive to systematics.
This criterion reduces the sample by 20 \%, as can be inferred from table~\ref{tab:targets}. 
Figure~\ref{fig:mag_cut} shows that the correlation function at large $r$ is closer to zero after this cut. 
We obtain $ \langle {r.\;\xi(r)} \rangle = 0.31 \pm 0.10$ over the range $120 < r < 1500$ $h^{-1}$Mpc.

\begin{figure}[htb]
\begin{center}
\epsfig{figure=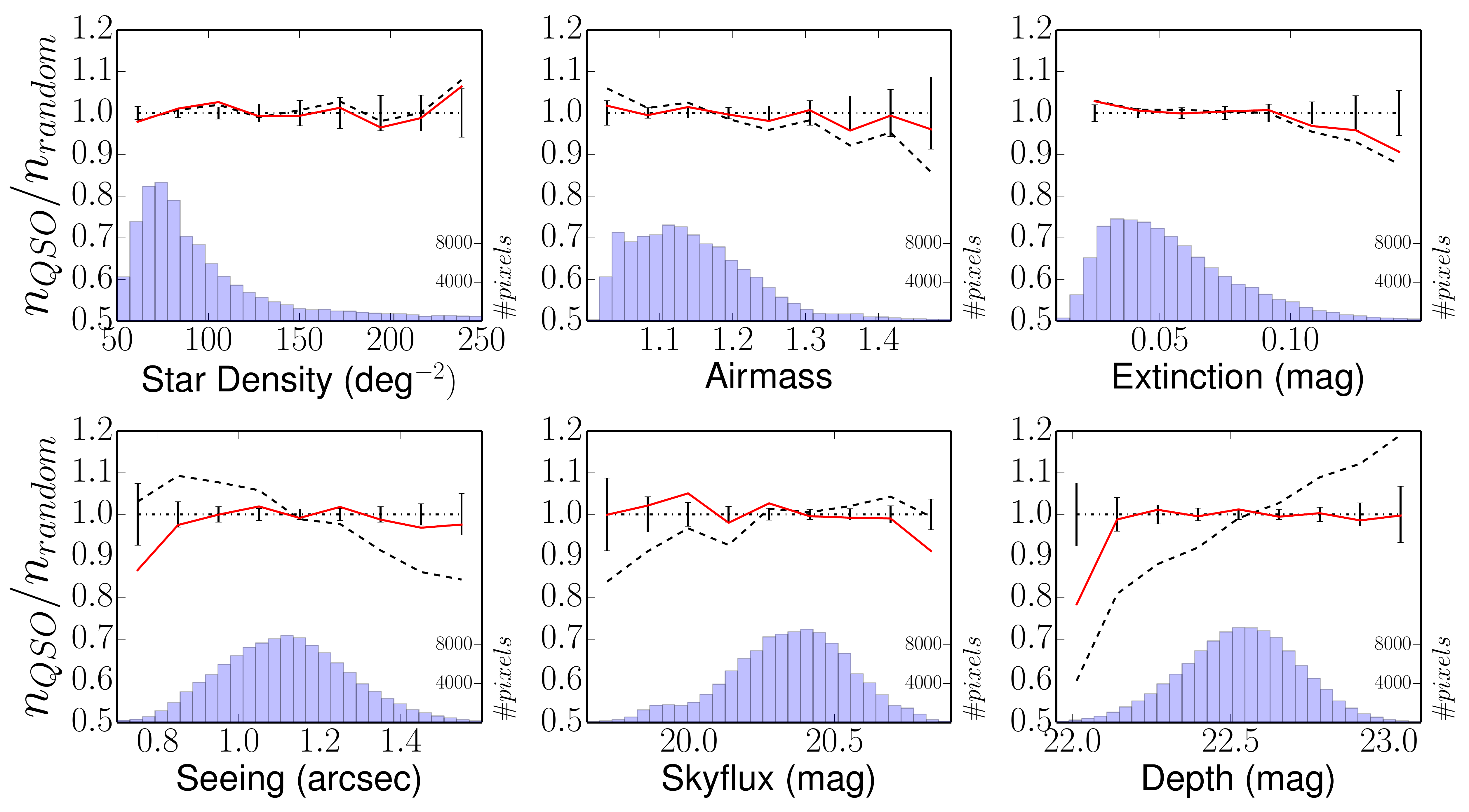,width = 15cm} 
\caption{\it The $n_{QSO} / n_{random}$ ratio versus star density, airmass, Galactic extinction, seeing, sky brightness (magnitude corresponding to the sky flux in 1 arcsec$^2$) and survey depth (i.e.,\ 5$\sigma$ detection limit) in the $i$-band before (black dotted line) and after weighting depending on survey depth (continuous red line). The error bars are practically the same for the two curves. For sake of clarity the errors are drawn only once, on the line $n_{QSO} / n_{random}=1$. The histograms show the distribution of the  number of Healpix pixels for each quantity.}  
\label{fig:dependencies}
\end{center}
\end{figure}

With this apparent magnitude cut set, we perform some simple checks on the analysis. We split our quasar sample in two subsamples of the same size, including  quasars with seeing below or above the median seeing value in the $r$-band. The correlation functions for the two subsamples are found to be  compatible. The same operation is performed with star density, airmass, Galactic extinction and sky brightness, leading to the same conclusion.
One also must be aware of a possible systematic shift between redshifts obtained using different emission lines. In our redshift range, the Mg\textsc{II} emission line is the most reliable line for fixing redshifts, although it is shifted to wavelengths that are beyond the BOSS coverage for redshifts above 2.5.
As an additional check, we select quasars whose redshift is assigned with the Mg\textsc{II} line and compute the correlation function for this subsample ; the results are compatible with the correlation function for all quasars in the $2.2-2.5$ redshift range.

Even after the apparent magnitude cut, targets remain closer to the detection limit in some parts of the sky, although with a safety margin.
We therefore study whether the measured quasar density presents a dependence with respect to the 5-$\sigma$ detection limit and to its inputs: the seeing, airmass, Galactic extinction and sky brightness. In addition, we consider the dependence with star density, since it influences the amplitude of blending effects. This procedure is similar to what was done in Ross et al.~\cite{RossA+12} for galaxy clustering.
A HEALPix~\cite{HEALPix} map ($N_{side} = 256$) is produced for each quantity using SDSS photometric data, and also for the ratio of the number of measured quasars to the normalized number of randoms. The black lines in figure~\ref{fig:dependencies} show that this ratio varies significantly with all quantities except star density, suggesting that there is actually not much blending effect for our quasar sample.

The dependence with respect to the detection limit is independent of redshift and can be fitted with a straight line.
This may be corrected by weighting the quasars in each pixel by the inverse of the value of the straight line fit corresponding to the 5-$\sigma$ detection limit in the considered pixel.
The red lines in figure~\ref{fig:dependencies} represent the new resulting $n_{\rm QSO}/n_{\rm random}$ ratios. 
The effects of systematics are then minimal. 
The blue points in figure~\ref{fig:mag_cut} demonstrate that after applying this weighting, the correlation function at large $r$ is consistent with zero. 
We find $ \langle {r.\xi(r)} \rangle = 0.09 \pm 0.07$ over the range $120 < r < 1500$ $h^{-1}$Mpc.

%--------------------------------------------------------------------------------------------------------------------
\subsection{Analysis-related effects}
\label{sec:D2bias}

In our determination of $\sNr$, we do not limit ourselves to spheres entirely included in the survey and we correct for the missing parts of the spheres by using random catalogues. In order to study whether this procedure can bias the measured $\sNr$ and $D_2(r)$, we generate fractal distributions over the survey geometry with a dimension below 3. We then send them through the analysis pipeline to check the reconstructed $\widehat D_2(r)$.

In a first step, following Castagnoli \& Provenzale ~\cite{Castagnoli+91}, we create a cube embedding the full survey, with side $L = 36$ $h^{-1}$ Gpc. We divide this cube in $M=8$ sub-cubes of side $L/m$ with $m=2$, and give to each sub-cube a survival probability $p$. We repeat this procedure for each surviving sub-cube, and so on. For an infinite number of iterations this provides a monofractal distribution with 

\begin{equation}
D_{2} = \frac{\log pM}{\log m} = 3 +\frac{\log p}{\log 2} \; .
\end{equation}
This formula shows that using $p \lesssim 1$ gives $D_2 \lesssim 3$.

In a second step objects are drawn in the final sub-boxes. 
The number of objects in each sub-box is drawn using a Poisson distribution of mean $\lambda < 1$ and the position of the objects are randomly drawn in the sub-box. $\widehat D_2(r)$ was computed for the resulting objects over the full (36 $h^{-1}$ Gpc$)^3$ box using only spheres entirely included in the box. 
Eleven iterations in the first step were found to be enough to provide $\widehat D_{2}(r)$ compatible with its expected value, even after averaging over 100 realizations of the full box.

\begin{figure}[t]
\begin{center}
\epsfig{figure=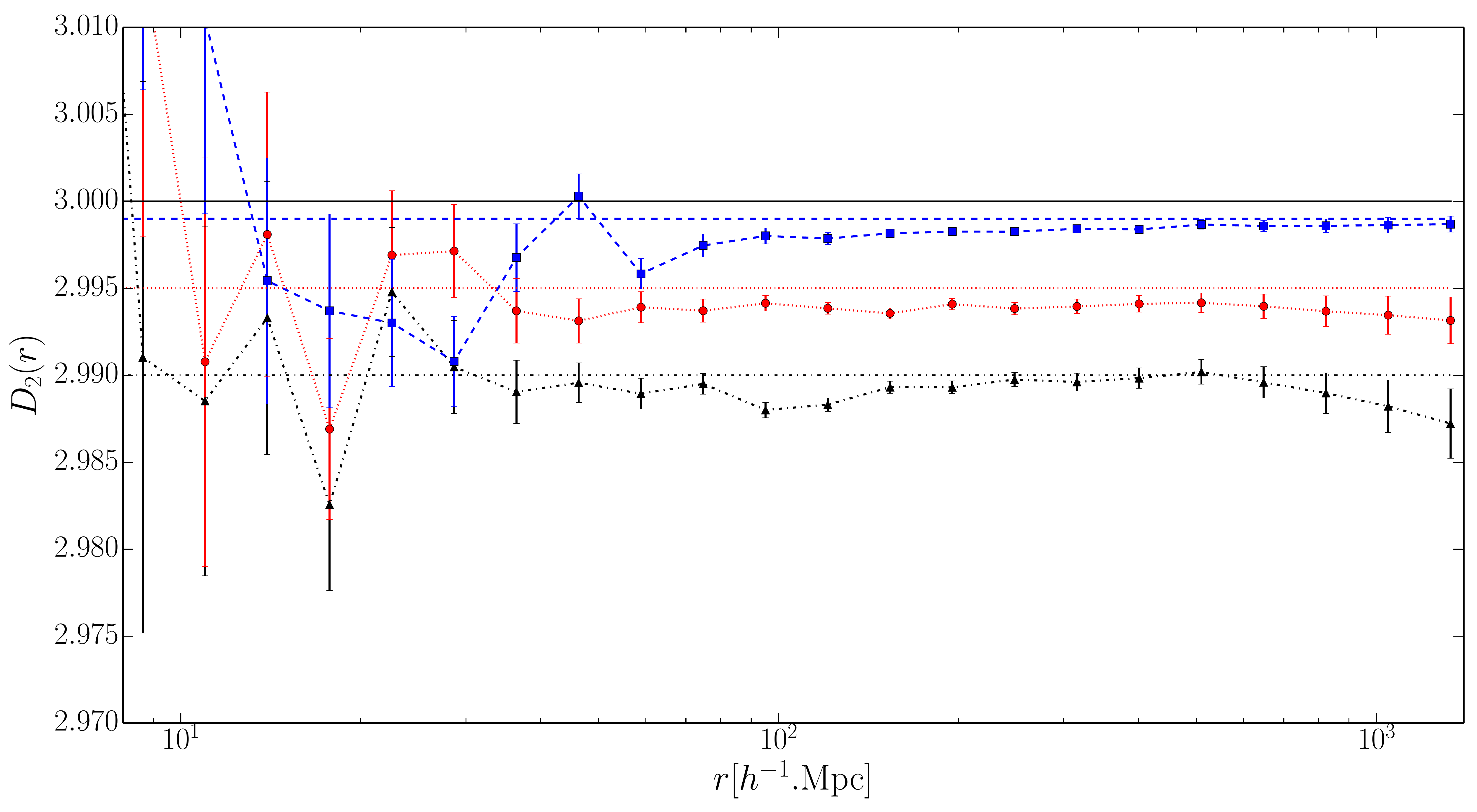,width = 12cm} 
\caption{\it Reconstructed $D_2(r)$ averaged over 100 realizations for fractal distributions of various dimensions. The horizontal dashed or dotted lines indicate the input value of $D_2$.}
\label{fig:fractal}
\end{center}
\end{figure}

Finally, the positions of the objects are converted from cartesian coordinates to observational coordinates RA, DEC and $z$. Cuts are applied to reproduce the footprint, masks, completeness and $n(z)$ of the survey. The mean value of the Poisson distribution for the number of objects per sub-box is set so that the total number of objects after cuts matches the number of quasars in the real data.
The resulting $\widehat D_{2}(r)$ averaged over 200 realizations are shown in figure~\ref{fig:fractal}. The errors are obtained from the spread of the realizations. The reconstructed $\widehat D_2(r)$ appear to be in reasonable agreement with the input. In particular, if there is a bias, it is an increase of the difference between $D_2$ and 3. So the upper limits on $3-D_2$ that we quote remain valid.

\newpage

%%%%%%%%%%%%%%%%%%%%%%%%%%%%%%%%%%%%%%%%%%%%%%%%%%%
\section{Results}
\label{sec:results}

%--------------------------------------------------------------------------------------------------
\subsection{Model-independent study} % towards a model independent study
\label{sec:DDRRresults}

\begin{figure}[t]
\begin{center}
\epsfig{figure= 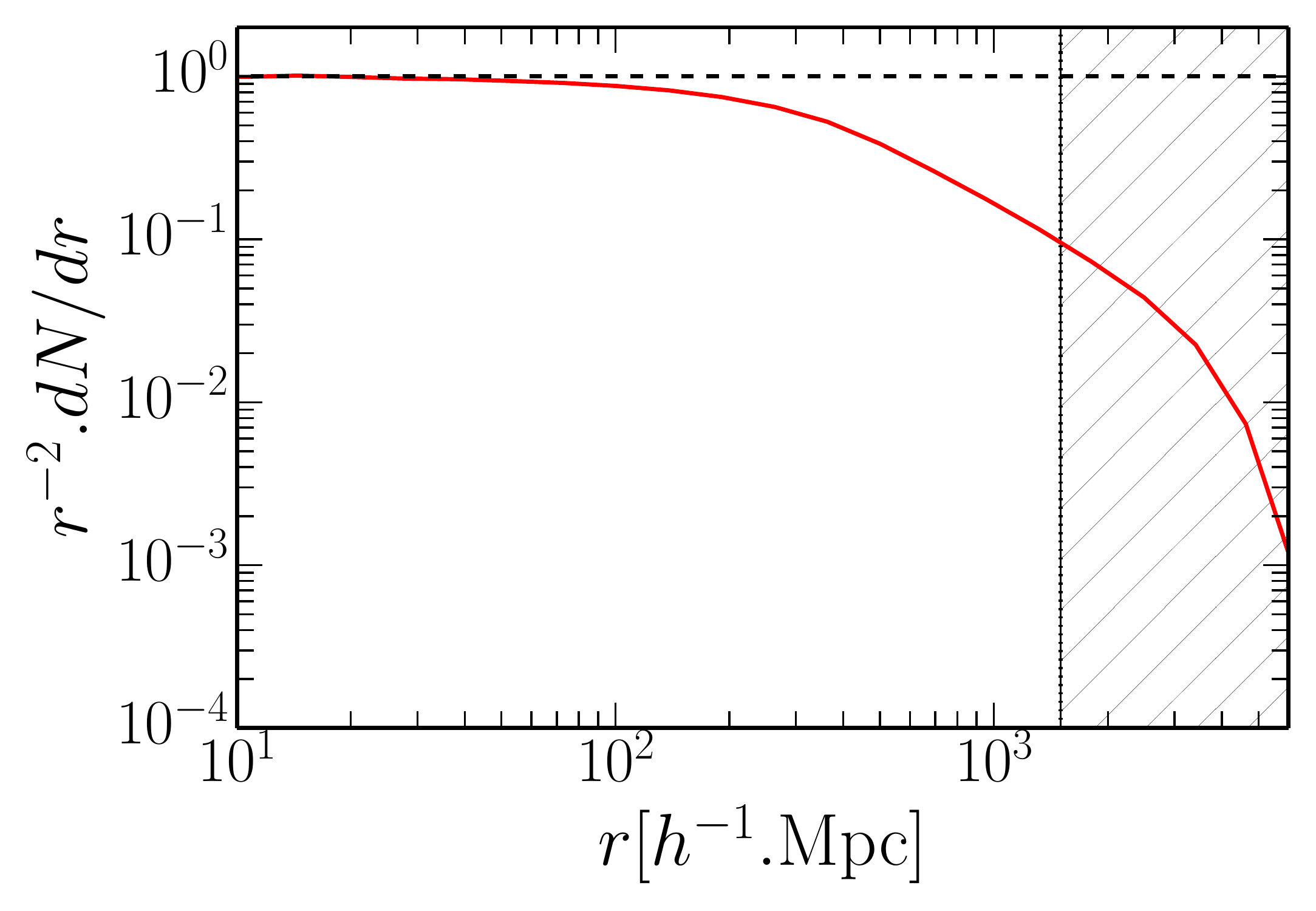,width = 8cm} 
\caption{\it Density of pairs from the random catalogue separated by a distance $r$, divided by $r^2$, 
$\displaystyle \frac{1}{r^2}\frac{dN}{dr}$,
and normalized such that it is unity at small $r$. 
At large $r$ this scaled density decreases due to the finite volume of the survey.
For $r>1500 h^{-1}$ Mpc, it gets too low, and the hatched area is not included in the analysis.
}
\label{fig:RR}
\end{center}
\end{figure}

Figure~\ref{fig:RR} shows the density, $dN/dr$, of pairs from the random catalogue separated by a distance $r$, divided by $r^2$. This scaled density is constant at small $r$ and then decreases due to the finite volume of the survey. 
At $r=1500$ $h^{-1}$Mpc it is $\approx$10\% of its maximal value. The actual number of pairs in a bin of $r$ is still increasing with $r$, so there is still information in $\Nr$ or $\sNr$. This is increasingly less the case for larger values of $r$ and we decide to stop the analysis at $r=1500$ $h^{-1}$Mpc for $\sNr$. For $D_2(r)$, which is a derivative of $\sNr$, the analysis goes up to $r=1200$ $h^{-1}$Mpc.

\begin{figure}[htb]
\begin{center}
\epsfig{figure= 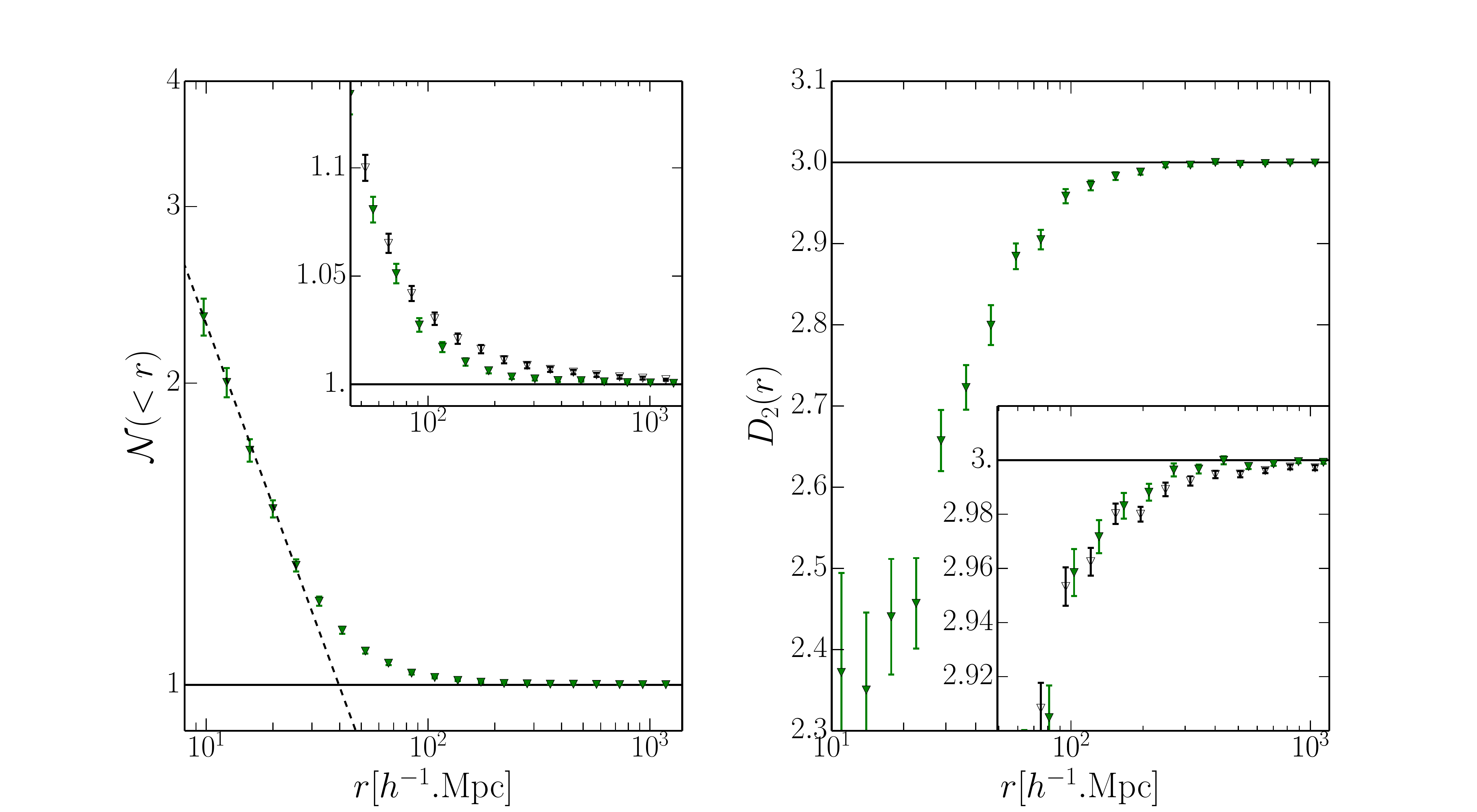,width = 16cm} 
\caption{\it  Scaled count-in-sphere $\sNr$ (left) and correlation dimension $D_2(r)$ (right) for the quasar sample as a function of the sphere radius $r$. These measurements are obtained applying the masks and the  magnitude cut (green triangles). The dashed line in the left panel is a power law fit for $6<r<25$ $h^{-1}$Mpc. 
Each insert displays, in addition, the same observable without the mask and the magnitude cut (black open triangles). 
A constant $\sNr$ and $D_2(r)$ compatible with 3 establish homogeneity on large scales.
}
\label{fig:resuQSO}
\end{center}
\end{figure}

The green points in figure~\ref{fig:resuQSO} are  $\sNrHat{LS}$  and $\widehat D_2(r)$ values obtained when applying the masks and the magnitude cut. It is safe to apply them since they are uncorrelated with the distribution of extragalactic quasars.
We did not apply, however, the weighting depending on photometric survey depth. This weighting implies fitting the data to make the density flat with respect to the depth and it may slightly bias the measurement towards homogeneity.
Due to low quasar density, the measurement does not extend to very low values of $r$ and the errors are significant at small $r$. 
For $6<r<25$ $h^{-1}$Mpc,  $\sNr$ is clearly compatible with a power law, $\sNr \propto r^{-\gamma}$ with $\gamma=0.60 \pm0.03$, corresponding to a fractal correlation dimension, $D_2=2.40\pm 0.03$.
This result contrasts with the large scale regime ($r \gtrsim 100$ $h^{-1}$Mpc) where $\sNr$ is found to be constant over one decade in $r$.

However, as illustrated in figure~\ref{fig:corrmatrix}, $\sNr$ data points are significantly correlated, making their interpretation difficult. It is therefore better to examine the  $D_2(r)$ data points, which are nearly uncorrelated. 
The correlation dimension $D_2(r)$  reaches 3 at large $r$. Taking into account the covariance matrix, we get $3 - \langle D_2 \rangle = (1.1 \pm 0.3) \times 10^{-3}$ (1 $\sigma$) over the range $250<r<1200$ $h^{-1}$Mpc. 
This is small but not compatible with zero. We ascribe that to the inhomogeneities in the target selection discussed in section~\ref{sec:syst}. Indeed, when we apply the depth-dependent weight in section~\ref{sec:LSresults}, $D_2$ gets compatible with 3. Therefore we present our result as a 2$\sigma$ upper limit, $3 - \langle D_2 \rangle < 1.7 \times 10^{-3}$, i.e.~$\langle D_2\rangle > 2.9983$.
This establishes the homogeneity of the quasar distribution to a high accuracy.

The black points in figure~\ref{fig:resuQSO} are  $\sNrHat{LS}$ and $\DHat{LS}$ values obtained without applying the masks and the apparent magnitude cut. Even in these conditions the data indicate homogeneity, although with less accuracy. This means that our conclusions on homogeneity are quite robust and do not depend on analysis subtleties.

The limit obtained on $3-D_2$ is for the quasar distribution. Strictly speaking this is the only thing that can be said in a model independent way. Indeed, the quasar bias is required in order to derive a limit for the matter distribution. 
In section~\ref{sec:bias} we obtain the quasar bias by comparing our measured quasar correlation function to the matter correlation function extrapolated (in redshift) from the measured amplitude of the CMB fluctuations. This extrapolation takes into account the growth factor between decoupling and $z=2.39$ in our $\Lambda$CDM model, it is then model dependent. 

Although this requires some (reasonable) assumptions, one can argue that the quasar bias at $z=2.39$ must be larger than unity.
Weak lensing is not sensitive to a peculiar tracer but to the total matter inhomogeneities and we can compare the clustering measured with weak lensing to the quasar clustering we measure. Using only weak lensing data, figure 2 in \citet{Bacon+05} shows that at $z \approx 0.2$ the clustering of matter is at most 1.6 times the $\Lambda$CDM prediction  (2$\sigma$).
 We measure an effective quasar bias of 4.25, which means that if we compare our measured quasar clustering amplitude to the $\Lambda$CDM prediction for clustering of matter at $z=0.2$, we get a factor $4.25^2 \times \left. D^2(z=2.39) \middle/ D^2(z=0.2)=3.05 \right.$, where $D(z)$ is the growth factor. % (4,25 x 0.370 / 0.9)^2.
 So we measure an amplitude of clustering that is at least 3.05/1.6=1.9 times that measured by \citet{Bacon+05}. We stress that the $\Lambda$CDM prediction for clustering of matter at $z=0.2$ cancels in this ratio, which is then just the ratio of two measurements.
If we assume that matter clustering does not decrease with time, then the bias of quasars at $z=2.39$ is at least $\sqrt{1.9}$. Comparing the clustering of different tracers with different biases shows that the bias ratio does not evolve much with the scale; we may then assume that the bias measured at scales smaller than typically 50 $h^{-1}$Mpc can be used to correct for $\sNr$ and $D_2$ on large scales.
Making these assumptions, we can set a limit on the order of  $D_2-3 < 0.9 \times 10^{-3}$ for matter distribution.

\citet{Labini11} studied cosmic homogeneity using the main galaxy catalogue of SDSS DR7~\cite{Abazajian+09}. In the range $5<r<20$ $h^{-1}$Mpc, he found a power law index $\gamma = 0.88 \pm 0.05$, i.e., $D_2=2.12\pm0.05$. He also fit a power law in the region $30<r<150$ $h^{-1}$Mpc, although this function does not perfectly fit his data, and found $\gamma = 0.2 \pm 0.05$, i.e., $D_2=2.8\pm0.05$. 
If we force a power law fit in this $30<r<150$ $h^{-1}$Mpc range to our data, we find $\gamma = 0.109 \pm 0.012$. This result is also significantly different from zero, which is not a surprise since this is the region where we observe a transition to homogeneity. The two sets of data qualitatively agree with each other: fits to the data produce power laws with different non-zero indices below $r=20$ $h^{-1}$Mpc and for $30<r<150$ $h^{-1}$Mpc, although the data are poor matches to the power law in the second range in $r$.
The small level of discrepancy might be ascribed to the different range in redshift. In addition the completeness of the survey and the masked regions are not taken into account in the \citet{Labini11} analysis. This should affect $\sNr$ at some level because the completeness and the masks are not randomly distributed, and not accounting for them produces fake correlations.
The important point is that the data used by \citet{Labini11} do not extend to the region $r>150$ $h^{-1}$Mpc, where we clearly observe homogeneity.

%--------------------------------------------------------------------------------------------------
\subsection{Quasar bias}
\label{sec:bias}

\begin{figure}[t]
\begin{center}
\epsfig{figure=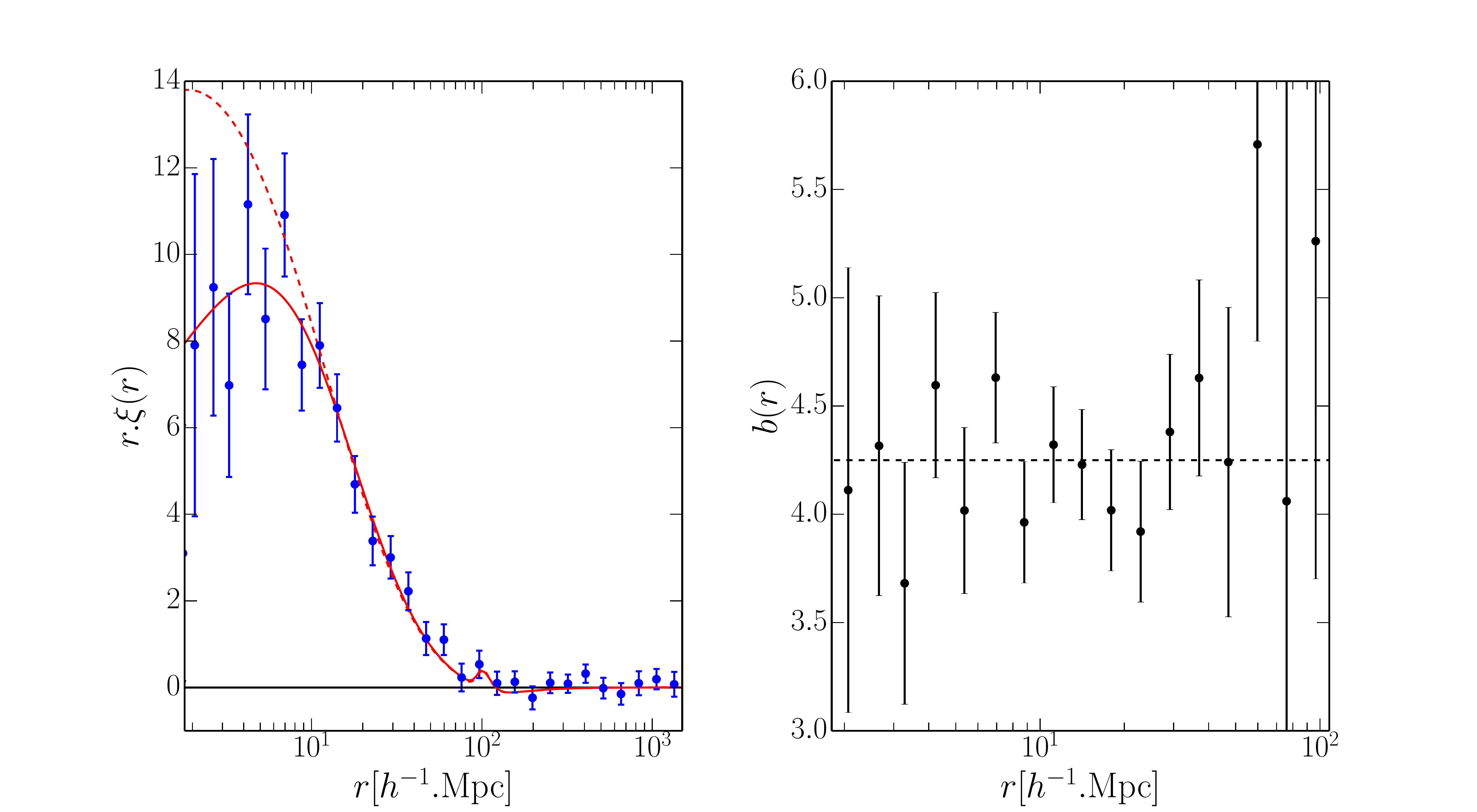 ,width = 16cm} 
\caption{\it Left: the two-point correlation function of quasars in the BOSS sample as a function of their separation in redshift space. The dotted curve is the CAMB linear prediction (fitted over $10<r<85$ $h^{-1}$Mpc) while the continuous curve is the nonlinear CAMB model with velocity dispersion (fitted over $2<r<85$ $h^{-1}$Mpc). The continuous curve has $\chi^2=9$ for 16 degrees of freedom.
The model is negative for $r>120$ $h^{-1}$Mpc but indistinguishable from zero given the error of our data. Right: resulting quasar bias as a function of the separation~$r$. The measured bias does not depend on the selected range in $r$.
}
\label{fig:xi}
\end{center}
\end{figure}

We fit our measured 2-point correlation function, $\xi(r)$, with the model described in section~\ref{section:model}, where we use $\Lambda$CDM prediction obtained with the CAMB program~\citep{cambref} for $P_m(k)$. 
As stated in the introduction we use the Planck 2013 + WMAP9 parameters~\cite{Planck13}.
The covariance of the data is obtained by bootstrap resampling, as discussed in section \ref{sec:bootstrap}. 
As illustrated in figure~\ref{fig:xi} (left), we use two versions of the model, a linear CAMB prediction (dotted curve) with velocity dispersion set to zero, and a non-linear CAMB version provided by HALOFIT~\cite{halofit} with velocity dispersion (solid curve). 
There is a clear difference for scales below 15 $h^{-1}$Mpc, while the two predictions are in close agreement on larger scales. 
This difference arises mainly from the velocity dispersion and not HALOFIT.
The non-linear fit with velocity dispersion perfectly describes the data. 
Fitting from $r=2$ $h^{-1}$Mpc up to $r=85$ $h^{-1}$Mpc, where the measured $\xi(r)$ becomes compatible with zero, produces a bias $b_Q=3.91\pm0.14$, and a pair-wise velocity dispersion, $\sigma_p=416\pm101$ $\mathrm{km.s^{-1}}$, with  $\chi^2=9$ for 16 degrees of freedom. 
Figure~\ref{fig:xi} (right) shows that the bias is quite independent of the scale. 
Indeed, limiting the range of the fit to e.g.~$2<r<22$ $h^{-1}$Mpc, hardly changes the result, $b_Q=3.85\pm0.19$.
In addition, fitting with linear CAMB without velocity dispersion, but limiting the range of the fit to $10<r<42$ $h^{-1}$Mpc yields the same bias, $b_Q = 3.87 \pm 0.14$.
Therefore our estimation of the bias appears quite robust. 
Finally, table~\ref{tab:b_z} shows the measured bias obtained in various redshift bins. 
One of the bins, $2.4<z<2.8$, has a relatively poor $\chi^2$ of 27 for 16 d.o.f., which has a probability of 4\%.

\begin{table}
\centering
\begin{tabular}{|cccccc|} \hline
$z_{min}$ & $z_{max}$ & $z_{eff}$ & $b_Q$ & $\sigma_p$ (km s$^{-1}$)& $\chi^2$ (16 d.o.f.) \\  \hline
2.2 & 2.4 & 2.30 & $ 4.01 \pm 0.22$ & $757 \pm 214$ & $8$  \\
2.4 & 2.8 & 2.56 & $ 3.86 \pm 0.19$ & $314 \pm 110$ & $27$  \\
2.8 & 3.5 & 3.09 & $ 5.06 \pm 0.83$ & $708 \pm 672$ & $16$  \\ \hline
2.2 & 2.8 & 2.39 & $ 3.91 \pm 0.14$ & $416 \pm 101$ & $9$ \\
\hline
\end{tabular}
\caption{Fit of bias over $2<r<85$ $h^{-1}$Mpc in various redshift bins.}
\label{tab:b_z}
\end{table}

The result of the fit for $\sigma_p$ is in agreement with the estimation, 300 $\mathrm{km.s^{-1}}$, obtained in section \ref{section:model}.
The value of the quasar bias is in agreement with the fit $b_Q=0.53 +0.289\;(1+z)^2$ by \citet{Croom+05}, which yields 3.85 at our average $z=2.39$.
There exist several other determinations of the quasar bias from BOSS data, namely $b_Q=3.8 \pm 0.3$ from DR9 data~\citep{white12}, $b_Q=3.54 \pm 0.11$ from DR12 data~\cite{Eftekharzadeh+15} 
and $b_Q=3.64 \pm 0.15$ from the cross correlation of quasars and the Lyman-$\alpha$ Forest in DR9~\citep{font13}.
Our value, $3.91 \pm 0.14$, is slightly higher on average,
but we include dispersion in our model of the correlation function, corresponding to eq.~\ref{Eq:Pmodel} for $P(k)$, while these previous studies did not.
They also limit the fit to a maximum $r\approx 25$ $h^{-1}$Mpc. If we set $\sigma_p=0$ in our model, we obtain $b_Q=3.66 \pm 0.10$. If in addition we limit the fit to the range $2<r<26$ $h^{-1}$Mpc, the result is $b_Q= 3.57 \pm 0.11$, in fair agreement with all other measurements from BOSS data. 

In order to transform the $\sNr$ data for quasars to $\sNr$ for matter, using eq.~\ref{Eq:debias}, we need the effective bias, which includes the effect of linear redshift-space distortions.
Fitting $\xiHat{LS}$ in the range $2<r<85$ $h^{-1}$Mpc with the model corresponding to eq.~\ref{Eq:Pmodel} with $b^2(1+\beta\mu^2)^2$ replaced by $b_{\rm eff}^2$, results in $b_{\rm eff} = 4.25 \pm 0.14$.
Alternatively,  $\langle z \rangle =2.39$ produces $\Omega_m = 0.947$, $f=0.970$ and $\beta =0.249$. 
Adopting the Kaiser formula ~\citep{kaiser87}, $b_{\rm eff}^2 = (1+2\beta/3+\beta^2/5)b^2$, with $b=3.91$ yields $b_{\rm eff}=4.24$.

%-------------------------------------------------------------------------------------------
\subsection{Cross check of $\Lambda$CDM model}
\label{sec:LSresults}

\begin{figure}[t]
\begin{center}
\epsfig{figure= 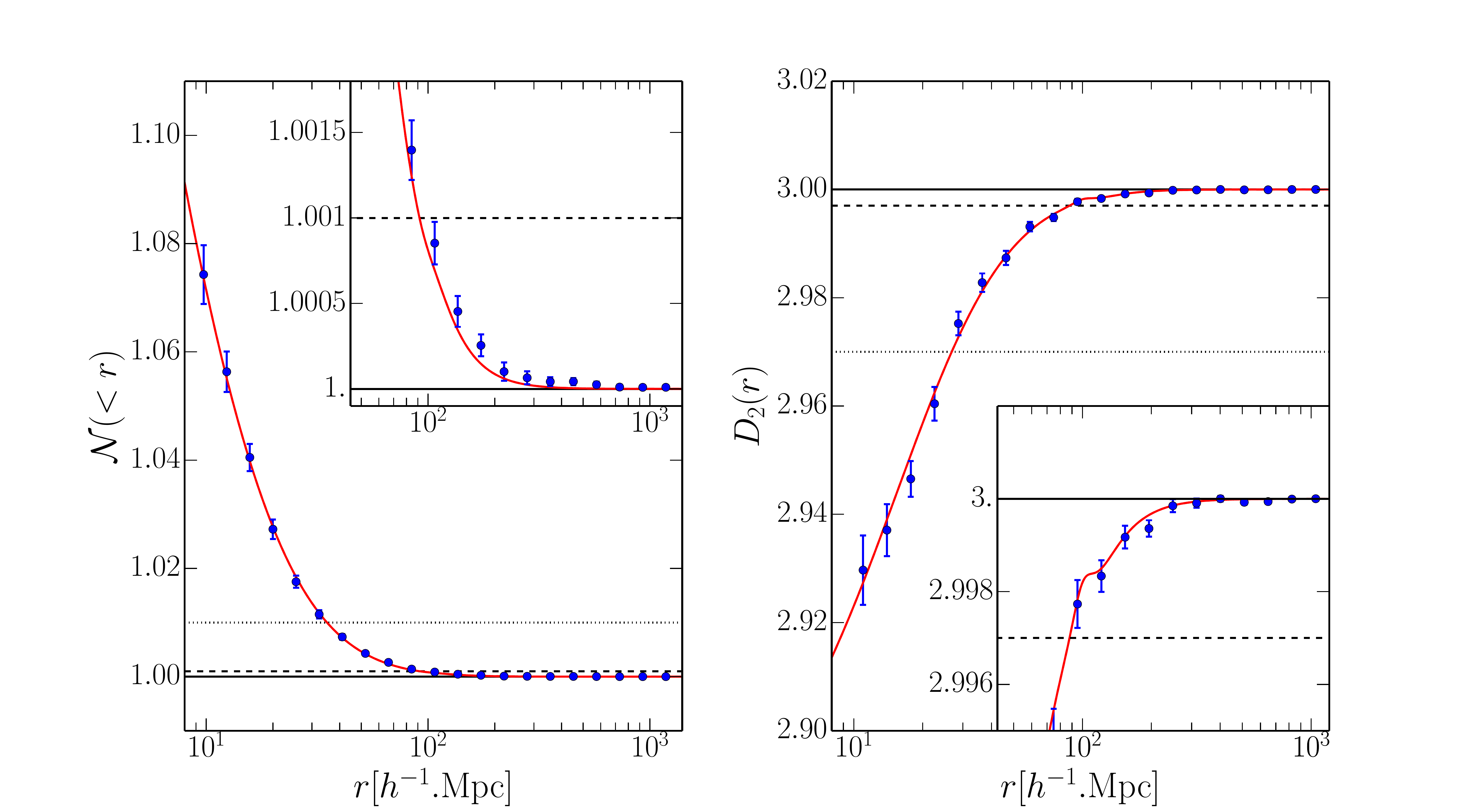,width = 15.5cm}
\caption{\it  Scaled count-in-sphere $\sNr$ (left) and correlation dimension $D_2(r)$ (right) for the matter distribution as a function of the radius of the sphere $r$. 
The red curve is the $\Lambda$CDM model and the wiggle at $r\approx$ 100 $h^{-1}$Mpc in $D_2(r)$ is the BAO peak. 
The horizontal lines indicate the limits that define the one percent (dotted lines) and one per mil (dashed lines) homogeneity scales. 
The transition to homogeneity is accurately described by the $\Lambda$CDM model with $\chi^2=16$ and 12 for 18 degrees of freedom for $\sNr$ and $D_2(r)$, respectively.
}  
\label{fig:Nr}
\end{center}
\end{figure}

\begin{figure}[t]
\begin{center}
\epsfig{figure= 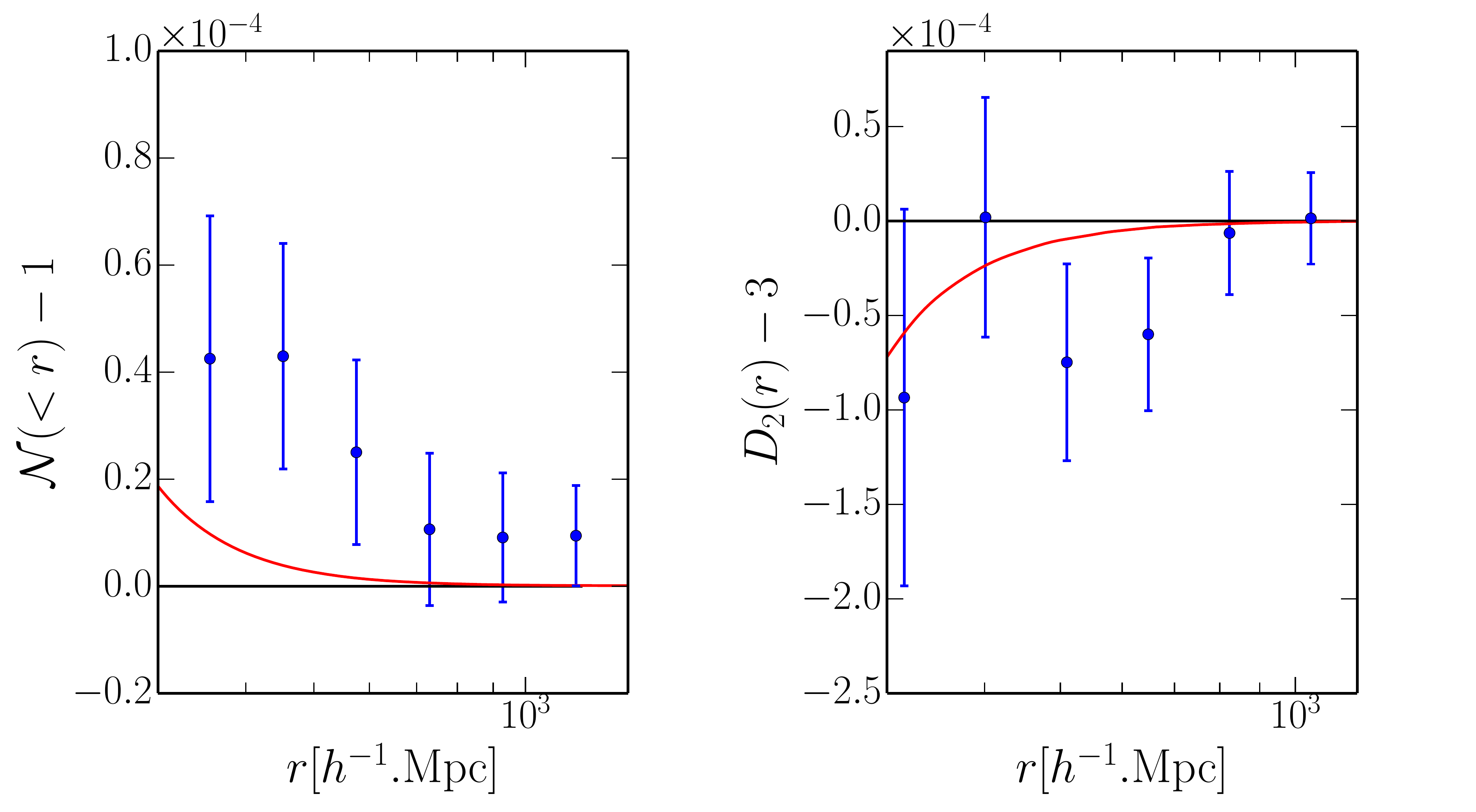 ,width = 11.5cm} 
\caption{\it  Expanded view of figure~\ref{fig:Nr} for $r>300$ $h^{-1}$Mpc:  $\sNr -1$ (left) and $D_2(r) -3$ (right). The red curve is the $\Lambda$CDM model.  In this range of $r$, the $\sNr$ points are highly correlated. 
The right panel shows that $D_2(r)$ is statistically compatible with 3 at a high accuracy (note the $10^{-4}$ factor applying to the vertical scale).
} 
\label{fig:Nrzoom}
\end{center}
\end{figure}

We now present an analysis that is no longer model independent, in contrast with section \ref{sec:DDRRresults}, but is an accurate cross check of $\Lambda$CDM model in terms of the transition to homogeneity.
We use the effective quasar bias relative to  $\Lambda$CDM model and eq.~\ref{Eq:debias}  to correct the measurements and obtain $\sNr$ for the matter distribution, and the resulting $D_2(r)$.
In addition to the masks and the magnitude cut we now apply the weighting depending on survey depth (section~\ref{sec:syst}).
We also apply eq.~\ref{Eq:debias} to the model of section~\ref{section:model} to produce the corresponding $\Lambda$CDM predictions.
Results are shown in Figure~\ref{fig:Nr} and  the region $r>250$ $h^{-1}$Mpc is expanded in Figure~\ref{fig:Nrzoom}.
The transition to homogeneity is clearly visible, and the behaviors of $\sNr$ and $D_2(r)$ are accurately reproduced by the $\Lambda$CDM model (red curve). For $\sNr$ several adjacent points in Figure~\ref{fig:Nrzoom} are above the curve, but the points are highly correlated in this region of $r$, as can be seen in figure~\ref{fig:corrmatrix}. Indeed, when the full covariance is taken into account, we find $\chi^2=16$ for 18 degrees of freedom over the range $6<r<1200$ $h^{-1}$ Mpc. A similar $\chi^2$ is found for $D_2(r)$, 12 for 18 degrees of freedom. 
In the region $250<r<1200$ $h^{-1}$Mpc, we have for the matter distribution $3-\langle D_2 \rangle = (1.8 \pm 1.7) \times 10^{-5}$ (1 $\sigma$), which gives an upper limit $3-\langle D_2 \rangle < 5.2 \times 10^{-5}$ (2 $\sigma$).  We note that with the weighting, $D_2(r)$ is now compatible with 3, in contrast to what was shown in section~\ref{sec:DDRRresults}.

\citet{Scrimgeour+12} introduce two homogeneity scales, $R_H$. They are defined as the values of $r$ for which $\sNr$ and $D_2(r)$ reach their nominal value within one percent, namely 1.01 for $\sNr$ and 2.97 for $D_2(r)$ ; they also define the values for 1 per mil. These thresholds are indicated in figure~\ref{fig:Nr}. 
Table~\ref{tab:R_H} shows that our measurements of $R_H$ are in good agreement with the values in our $\Lambda$CDM fiducial model.

However, our measurements of $R_H$ depend on the value of the bias, which is obtained from the same data.
Therefore, for $R_H$ at 1 percent, the agreement is of moderate interest. Indeed, in this case $R_H <$ 40 $h^{-1}$ Mpc, which is in the range of $r$ where the bias is obtained by fitting $\xi(r)$. So the agreement just means that $\xi(r)$ can be fitted with a $\Lambda$CDM model 
up to 40 $h^{-1}$Mpc. The agreement at 1 per mil is more interesting. 
The bias is essentially obtained from the range $0 < r < 40$ $h^{-1}$ Mpc, higher $r$ hardly contribute (see section~\ref{sec:bias}). So the agreement means that using a bias measured on small scales, $\Lambda$CDM gives a good description of the transition to homogeneity around 100 $h^{-1}$ Mpc.

In any case, the important result is the agreement of $\sNr$ and $D_2(r)$ with the fiducial $\Lambda$CDM model over a large range in $r$ with only the bias as a free parameter (we still obtain a good agreement if we use the value of $\sigma_p$ obtained in section~\ref{section:model} instead of fitting it to the data, and, in any case, $\sigma_p$ is irrelevant for $r>20$ $h^{-1}$Mpc for $\sNr$ and 30 $h^{-1}$Mpc for $D_2(r)$).

\begin{table}
\centering
\begin{tabular}{|l|cc|cc|} \hline 
				& \multicolumn{2}{|c|}{1 percent}		& \multicolumn{2}{|c|}{1 per mil} \\
				& $\sNr$ 			& $D_2(r)$		&  $\sNr$ 			& $D_2(r)$	\\ \hline
data				& $34.6\pm1.2$	& $26.2\pm0.9$	& $100\pm7$		& $88\pm5$	\\
$\Lambda$CDM	&	35.1			&	26.8			&	91.6			&	87.8		\\ \hline
\end{tabular}
\caption{The homogeneity scale $R_H$ in $h^{-1}$ Mpc for $\sNr$ and $D_2(r)$ at 1 percent and 1 per mil.}
\label{tab:R_H}
\end{table}

\citet{Yadav+10} define the homogeneity scale as the value of $r$ for
which the difference $3 - D_2(r)$ becomes smaller than the error on
$D_2(r)$. They derive an upper limit of $\approx 260$ $h^{-1}$ Mpc for this
homogeneity scale. This is quite consistent with our determination of
$D_2(r)$ in figure~\ref{fig:Nr}, right. However, the upper limit is valid only in the limit
of negligible shot noise, which is far from being our case. So this might
just be a numerical coincidence.
 
%%%%%%%%%%%%%%%%%%%%%%%%%%%%%%%%%%%%%%%%%%%%%%%%%%%%%%%%%%%%%%%%%%%%%%%%%%%%%
\section{Summary, discussion and conclusions}
\label{sec:discussion}

The BOSS quasar sample in the range $2.2<z<2.8$ was used to investigate cosmic homogeneity. The analysis was limited to 5983 deg$^2$ in order to mitigate systematic effects. These criteria provide a volume of 14 $h^{-3}$ Gpc$^3$ and allow a study of homogeneity up to a scale above 1 $h^{-1}$Gpc.

In order to mitigate systematic effects, we first apply masks based on Galactic objects and data acquisition, which is safe since they are uncorrelated with the distribution of extragalactic quasars. Faint quasars are more prone to systematics, indeed, they exhibit a different correlation function on large scales. We excise them, tightening the apparent magnitude cut until the correlation function is stable.  

We do not use the correlation function to study homogeneity because its definition depends on the average density, which is only defined for a homogeneous sample. We consider the counts-in-sphere $\Nr$, which is the average number of quasars in a sphere or radius $r$ around a given quasar, and the fractal correlation dimension, which is its logarithmic derivative, $D_2(r) = d \ln \Nr / d \ln r$. For a homogeneous sample $\Nr \propto r^3$ and $D_2 = 3$. 
We use a new estimator for $\Nr$ and $D_2(r)$, inspired from the Landy-Szalay estimator of the correlation function,
and that is more accurate than previous estimators by up to a factor 10 on the largest scales. 
We find $3-\langle D_2 \rangle <  1.7 \times 10^{-3}$ (2 $\sigma$) over the range $250<r<1200$ $h^{-1}$Mpc for the quasar distribution. 

However, we use a random catalogue to take into account the effect of survey geometry and completeness. 
The redshift distribution of this catalogue is taken from the data. This means that we are insensitive to a possible isotropic variation of density with redshift, $\rho=\rho(z)$. 
Our data cannot exclude, for instance, models where we live at the center of a spherical void.
We can test homogeneity up to a redshift dependence, or, in other words, check for spatial isotropy, i.e.~$\rho(r,\theta_1,\phi_1)=\rho(r,\theta_2,\phi_2)$ for any $(r,\theta_1, \phi_1,\theta_2,\phi_2)$.
Although this issue is not much discussed in 3D survey publications, this is actually happening for any 3D survey of a given type of source. One cannot disentangle source evolution from variations of matter density with redshift. Indeed, \citet{Mustapha+98}  demonstrated that 
{\it given any spherically symmetric geometry and any set of observations, we can find evolution functions that will make the model compatible with the observations}. % p 818

We can actually demonstrate this spatial isotropy without using any fiducial cosmology. 
Indeed, our analysis concludes that $\rho(\theta,\phi,z)$ = constant on large scales and this quantity can be obtained by applying a Jacobian $J=H(z)/c D_A^2(z)$ to the observed density of sources as a function of angle and redshift, $dN(\theta,\phi,z)/d\Omega dz$. 
So the result of our analysis is $\rho=J(z) \times dN(\theta,\phi,z)/d\Omega dz =$ constant on large scales. This means that $dN(\theta,\phi,z)/d\Omega dz$ depends only on $z$ on large scales, which is spatial isotropy.

In the framework of the Copernican principle, spatial isotropy implies homogeneity and therefore the cosmological principle.
This is not just using one principle to prove another principle because the two principles do not have the same status. 
In the context of this paper, the Copernican principle means that if the Universe exhibits spatial isotropy on large scales for us, it also exhibits it for other observers in the Universe. The opposite, i.e.~the idea that we are located in a kind of center of the Universe, would sound unreasonable to most physicists. On the other hand, the cosmological principle was mostly put forward to allow for a mathematical treatment of the problem. 

In addition, the Copernican principle itself was tested at the Gpc scale using spectral distortions of the CMB spectrum \cite{Stebbins07,CaldwellStebbins08} and the kinetic Sunyaev-Zel'dovich effect \cite{ZhangStebbins11}. 
These analyses rule out the adiabatic void model, where we live in the center of a large ($\sim$ Gpc) void, as an alternative to the existence of dark energy.

Previous analyses by the SDSS II and WiggleZ collaborations also implied a homogeneous Universe, or rather a spatially isotropic Universe. The SDSS II analysis was limited to $r \approx 100$ $h^{-1}$Mpc and only saw the onset of homogeneity. The WiggleZ analysis reached $\approx 300$ $h^{-1}$Mpc. 
An analysis of the SDSS II quasar sample in the range $1<z<1.8$ was recently published. It reaches $\approx 500$ $h^{-1}$Mpc and confirms homogeneity.
Our analysis in the range $2.2<z<2.8$ extends up to 1500 $h^{-1}$Mpc, which allows for unambiguously establishing homogeneity.
This result is at variance with another analysis of SDSS II data~\citep{Labini11}, but that analysis was limited to $r<150$ $h^{-1}$Mpc, so that it could only examine the transition region but not the large $r$ region where homogeneity is clear. 

Once homogeneity is established, we go beyond model-dependent analysis in order to quantitatively study the transition to homogeneity and perform a cross check of $\Lambda$CDM model.
we design a weighting scheme, based on the idea that average quasar density should not depend on the depth of the photometric survey.
This weighting results in a correlation function that is compatible with zero for separations larger than 100 $h^{-1}$Mpc. \citet{Karagiannis+14}  also measured the correlation function of the BOSS quasar sample on large scales, reporting it to be non-zero, and interpreted this result as an indication of primordial non-Gaussianities, which we do not confirm.
We ascribe the discrepancy to the fact that they used the DR9 sample, a period during which the definition of the CORE sample was not stable,  
and they did not apply the same cuts and weighting as we did to mitigate the effects of systematics.

We measure a quasar bias $b_Q=3.91\pm0.14$, relative to the $\Lambda$CDM model prediction for the matter correlation function. This result appears quite robust, and is in particular insensitive to the range of the fit. 
We have shown that small discrepancies between our result and earlier determinations of the quasar bias from BOSS data~\cite{white12,font13,Eftekharzadeh+15} result simply from model choices.

The resulting $\sNr$ and $D_2(r)$ for the matter distribution are found to agree with the predictions of $\Lambda$CDM, which therefore properly predicts the transition to homogeneity.
The fractal correlation dimension $D_2$ for the matter distribution is found to be compatible with 3 at a high accuracy:
$\langle 3-D_2 \rangle < 5.2 \times 10^{-5}$ (2 $\sigma$) over the range $250<r<1200$ $h^{-1}$Mpc. 

In conclusion, CMB, X-ray background and radiogalaxy measurements only demonstrate isotropy at a given redshift, or isotropy projected over a redshift range. 
Our data demonstrate spatial isotropy of the distribution of quasars on large scales, i.e.~at any redshift in the considered redshift range, $2.2<z<2.8$. This result is model independent, in particular it does not require a $\Lambda$CDM fiducial cosmology. 
Comparing quasar clustering to matter clustering measured by weak lensing, there are reasonable arguments that quasars do have a bias larger than unity independent of $\Lambda$CDM predictions. If we make this assumption, we get spatial isotropy of the matter distribution on large scales.
Combined with the Copernican principle it proves homogeneity on large scales over the redshift range $2.2<z<2.8$. Analyses of SDSS II (Galaxies and quasars) and WiggleZ data cover most of the lower redshift range. % WiggleZ 0.1<z<0.9  SDSSII QSO 1<z<1.8
So 3D surveys globally demonstrate homogeneity up to $z=2.8$.

On the other hand, if we use a $\Lambda$CDM fiducial cosmology and the quasar bias relative to the $\Lambda$CDM prediction, we get an accurate consistency check of  $\Lambda$CDM in terms of the transition to homogeneity on large scales

%%%%%%%%%%%%%%%%%%%%%%%%%%%%%%%%%%%%%%%%%%%%%%%%%%%%%%%%%%%%%%%%%%%%%%%%%%%%%
\appendix \section{New estimator of $\sNr$}
\label{add:LS}

If $\tilde N(s)$ is the number of pairs for a homogeneous sample, the Landy Szalay estimator of the correlation function is
\begin{equation}
1+\hat\xi^{\rm LS}(s) =\\ \frac{N(s)}{\tilde N(s)} = 1 + \frac{DD(s)-2DR(s)+RR(s)}{RR(s)}
\end{equation}

and
\begin{equation}
\displaystyle \sNr =\frac{\Nr}{\tilde N(\infr)} = \frac{\int_0^r \frac{N(s)}{\tilde N(s)}  \tilde N(s)\d s}{\int_0^r\tilde N(s)\d s} =
1 + \frac{\int_0^r [DD(s)-2DR(s)+RR(s)]\d s}{\int_0^r RR(s)\d s} \; ,
\end{equation}
where we used $RR(s)$ as an estimator of $\tilde N(s)$.

%%%%%%%%%%%%%%%%%%%%%%%%%%%%%%%%%%%%%%%%%%%%%%%%%%%%%%%%%%%%%%%%%%%%%%%%%%%%%
\acknowledgments
 We thank Jean-Philippe Uzan, Cyril Pitrou, Ruth Durrer and Martin Kilbinger for useful discussions. 
 We also thank the JCAP referee for very relevant comments. 

Funding for SDSS-III has been provided by the Alfred P. Sloan Foundation, the Participating Institutions, the National Science Foundation, and the U.S. Department of Energy Office of Science. The SDSS-III web site is http://www.sdss3.org/.

SDSS-III is managed by the Astrophysical Research Consortium for the Participating Institutions of the SDSS-III Collaboration including the University of Arizona, the Brazilian Participation Group, Brookhaven National Laboratory, Carnegie Mellon University, University of Florida, the French Participation Group, the German Participation Group, Harvard University, the Instituto de Astrofisica de Canarias, the Michigan State/Notre Dame/JINA Participation Group, Johns Hopkins University, Lawrence Berkeley National Laboratory, Max Planck Institute for Astrophysics, Max Planck Institute for Extraterrestrial Physics, New Mexico State University, New York University, Ohio State University, Pennsylvania State University, University of Portsmouth, Princeton University, the Spanish Participation Group, University of Tokyo, University of Utah, Vanderbilt University, University of Virginia, University of Washington, and Yale University.

\bibliographystyle{unsrtnat_arxiv}
\bibliography{biblio}	

\end{document}